\documentclass[fleqn,usenatbib]{mnras}

\usepackage{newtxtext,newtxmath}
\usepackage[T1]{fontenc}
\usepackage{ae,aecompl}
\usepackage{graphicx}
\usepackage{amsmath}
\usepackage{lipsum}

\usepackage{multirow}
\usepackage[section]{placeins}
\usepackage{mathrsfs}
\usepackage{ulem}
\usepackage[dvipsnames]{xcolor}
\usepackage{array, makecell} 
\usepackage{subcaption}
\usepackage{hhline}
\usepackage{blkarray}
\usepackage{libertine}

\newcommand\lsim{\mathrel{\rlap{\lower4pt\hbox{\hskip1pt$\sim$}}
        \raise1pt\hbox{$<$}}}
\newcommand\gsim{\mathrel{\rlap{\lower4pt\hbox{\hskip1pt$\sim$}}
        \raise1pt\hbox{$>$}}}

\title[The Z-dependent microphysics of stellar radii]{Dissecting the microphysics behind the metallicity-dependence of massive stars radii}

\author[Xin C., Renzo, M., \& Metzger, B.\ D.]{Chengcheng~Xin,$^{1}$
Mathieu Renzo,$^{2,1}$
Brian D.~Metzger,$^{1,2}$\\
$^{1}$Department of Astronomy, Columbia University, New York, NY,
10027, USA\\
$^{2}$Center for Computational Astrophysics, Flatiron Institute, New York, NY 10010, USA
}

\date{Accepted XXX. Received YYY; in original form ZZZ}

\pubyear{2022}

\begin{document}
\label{firstpage}
\pagerange{\pageref{firstpage}--\pageref{lastpage}}
\maketitle

\begin{abstract}
  Understanding the radii of massive stars throughout their evolution
  is important to answering numerous questions about stellar physics,
  from binary interactions on the main sequence to the pre-supernova
  radii. One important factor determining a star's radius is the
  fraction of its mass in elements heavier than Helium (metallicity,
  $Z$). However, the metallicity enters stellar evolution through
  several distinct microphysical processes, and which dominates can
  change throughout stellar evolution and with the overall magnitude
  of $Z$. We perform a series of numerical experiments with
  15$M_{\odot}$ \texttt{MESA} models computed doubling separately the metallicity
    entering the radiative opacity, the equation of state, and the
    nuclear reaction network to isolate the impact of each on stellar radii. We
  explore separately models centered around two metallicity values:
  one near solar $Z= 0.02$ and another sub-solar $Z\sim10^{-3}$,
    and consider several key epochs from the end of the main sequence
    to core carbon depletion.  We find that the
  metallicity entering the opacity dominates at most epochs for the
  solar metallicity models, contributing to on average $\sim60-90$\%
  of the total change in stellar radius. Nuclear reactions have a
  larger impact ($\sim50-70$\%) during most epochs in the subsolar $Z$
  models. The methodology introduced here can be employed more
  generally to propagate known microphysics errors into uncertainties
  on macrophysical observables including stellar radii.

\end{abstract}


\section{Introduction}
\label{sec:intro}

Stars are the most numerous observable objects in galaxies,
and much of astrophysics is pinned to the detailed
understanding of stellar physics. In particular, many applications
depend on precise knowledge of stellar radii.

Physically, stars extend into their winds without a well defined
end-point \citep[e.g.,][]{parker:58}, and observationally their radius
is a wavelength-dependent quantity because of the wavelength
dependence of the optical depth $\tau$. Nevertheless, the bulk of a
star's mass is contained within its photosphere\footnote{The
  photosphere is also wavelength-dependent. Here, we follow the
  standard assumption of defining the photosphere of a star where the
  optical depth is $\tau=2/3$ using the Rosseland mean opacity to
  calculate $\tau$.}.

The value of the photospheric radius $R$ enters in the global
dynamical timescale of the star
$\tau_\mathrm{dyn}\sim{}1/\sqrt{G\bar{\rho}}\sim{}\sqrt{R^3/GM}$,
where $G$ is the gravitational constant,  $\bar{\rho}$ is the average stellar density, expressed as
$\bar{\rho}\sim{}M/R^3$, and $M$ is the stellar mass. Therefore, the interpretation of dynamical phenomena, such as
pulsations, is sensitive to the precise value of the radius.  The
value of $R$ also enters directly in the surface gravity of a star,
thus in the interpretation of observed spectra. Whether or not a star
will interact with binary companion(s) depends on the extent of its
maximum radius \citep[e.g.,][]{sana:12}. If binary interactions occur,
the rate of radial change with mass ($\zeta=d\log R/dM$) contributes
to determining the dynamical stability of the mass transfer process
\citep[e.g.,][]{soberman:97, vigna-gomez:20}. Knowledge of the stellar
radius is often a limiting factor in the determination of the masses
and radii in eclipsing planetary systems (e.g., \citealt{Johnson+17}).
In the case of massive ($M\gtrsim 8\,M_\odot$) stars, their radial
extent at the end of their nuclear burning evolution sets the delay
between their final core collapse and shock breakout, or, in other
words, the delay between a possible neutrino and gravitational-wave
signal \citep[e.g.,][]{ott:09} and the first photons from a
(hypothetical) supernova explosion \citep[e.g.,][]{gill:22}.  Tidal interactions in stellar binaries are extremely sensitive to stellar radius, $\propto R^{\xi}$ where
$\xi\sim8$ (e.g., \citealt{Zahn77}).  Finally, stellar radii are also important in determining the
maximum luminosity achieved by individual stars, and thus by a given
stellar population, with cosmological implications
\citep[e.g.,][]{Jang+2017}.

In stellar evolution models of single stars, the radius depends
sensitively on many ``macrophysical'' and ``microphysical'' effects
described by parametric algorithms. These can impact stellar radii in
different ways \citep[e.g.,][]{Farrell+21}, and include among the
``macrophysical'' parameters those regulating mixing in the
stellar interior ($\alpha_{\rm MLT}$, overshooting, etc., see for
example \citealt{Dessart+2013}), wind (e.g., \citealt{renzo:17}) and
other mass loss mechanisms \citep[e.g.,][]{quataert:12, fuller:17,
  fuller:18, morozova:18}, rotation \citep[e.g.,][]{heger:00}, and the
primordial initial composition.

The primordial composition is particularly important, because
of the existence of observational tests comparing stellar populations
in M31, various parts of the Galaxy, the Magellanic Clouds, and other
dwarf galaxies
\citep{Bellazzini2001,Aloisi2007,Annibali2018,Wang2020,Martins2021}.
Moreover, the kind of binary interactions and the outcome of massive
binary evolution, especially for rare paths leading to the formation
of gravitational wave sources, are sensitive to the initial
composition \citep[e.g.,][]{Klencki+21,
  klencki:21b}.
The composition is usually summarized in terms of fraction of mass
made of helium ($Y$) and elements heavier than helium, so-called
metallicity $Z$, assuming a known scaling of the relative proportion
of each metal \citep[e.g.,][]{asplund:09, grasha:21}.

The composition also enters predominantly in determining the
``microphysical'' influence on the stellar radii. These can roughly be
divided in three categories relating to the radiative opacity
($\kappa$), the specific nuclear energy generation ($\epsilon$), and
the mean molecular weight ($\mu$) in the equation of state (EOS).

More specifically, the radiative opacity governs the energy transport
in radiative regions, such as the envelopes of main sequence (MS) massive
stars.  Indirectly, it also governs the temperature gradient and
thus the onset of convection, like in the core of MS massive
stars. Finally, the radiative opacity is key in the driving of massive
stars wind mass loss. The
dependence of $\kappa$ on composition comes dominantly from the bound-bound
and bound-free transitions which are the main source of opacity in the
envelope (conversely electron-scattering dominates in the
fully-ionized stellar core, \citealt{Kippenhahn&Weigert94}).

In massive MS stars with non-zero $Z$, the main source of nuclear energy
generation is the CNO cycle \citep[e.g.,][]{Bethe39}. In this cycle,
carbon (C), nitrogen (N), and oxygen (O) catalyze the conversion of
hydrogen (H) into helium (He), and the efficiency of the process
depends on the total mass fraction of these three elements. The
subsequent evolution is less directly dependent on the chemical
composition, as the mass fractions in the core are determined by the ashes
of preceding nuclear burning phases. Thus the role of the composition
in the MS nuclear burning governs the response of the
energy generation in the stellar core to surface energy losses via
photons (and later neutrinos, e.g., \citealt{fraley:68}).

Finally, the EOS determines the relation between thermodynamical
variables of the stellar gas, such as the pressure as a function of
density and temperature $P\equiv P(\rho, T, \{X_i\})$, where $\{X_i\}$ represents set of mass fractions for every isotope in the star.
In massive stars the
pressure is primarily due to non-degenerate gas, with a contribution
from radiation pressure the importance of which increases with stellar
mass. Typically, the total pressure is
  given by an ideal gas component, $P_\mathrm{gas}\simeq\rho T/\mu$, where
the composition enters through the mean molecular weight $\mu$,
plus a radiation pressure component independent of composition.
Since the hydrostatic equilibrium of a star is ensured by the balance
between gravity and the pressure gradient, the dependence of the
composition on $\mu$ impacts the radius indirectly.

Here, we isolate the metallicity-dependent effects of each
microphysical ingredient on the stellar radii. We perform a series of
controlled numerical experiments in which the metallicity used in each
microphysical ingredient can be varied separately. This allows us to
create unphysical stars with, for example, a different value of
metallicity for the radiative opacity and the nuclear burning or the
EOS. Our goal is to determine which microphysical input dominates at
each stage of evolution for different metallicities by comparing these
models.

This paper is organized as follows. We describe our \texttt{MESA}
stellar models and the numerical scheme we employ to dissect the microphysics-dependence of stellar radii in
section \ref{sec:method}.  After defining a quantitative measure of the fractional radius change due to each combination of microphysics input in section
\ref{sec:analysis}, we report our findings for select physical epochs of stellar evolution in the rest of section
\ref{sec:results}. In Sec.~\ref{sec:discussion}, we compare our results to previous studies investigating stellar radii from different
approaches, and explore the relative impact of particular choices made in our experiments (e.g., the assumed helium and hydrogen abundances; Sec.~\ref{sec:other_effects}).  Our
conclusions are summarized in Sec.~\ref{sec:conclusion}. In Appendix \ref{appendix:numerical} we quantify the main sources of numerical error which enter our radius estimates.

\section{Stellar models} \label{sec:method}

We compute stellar structure and evolution models using Modules for
Experiments in Stellar Astrophysics (\texttt{MESA}, version 12778;
\citealt{Paxton+2011,Paxton+2013,Paxton+2015,Paxton+2018,Paxton+2019}).
Each model consists of the evolution of a 15M$_{\odot}$ star from the
pre-main sequence to carbon depletion, defined as when the central
carbon mass fraction reaches $X_c({^{12}\mathrm{C}})\leq10^{-8}$.
Table~\ref{tab:models} lists all our models, and our input files and
numerical results are available at
\url{https://doi.org/10.5281/zenodo.6621643} or
\url{https://github.com/cx2204/stellar-radius}.

\begin{table*}
\centering
\begin{tabular}{c|l|c|c|c|c|c}
        & model \# (name) & $Z$ & $Z_{\mu}$ & $Z_{\kappa}$ & $Z_{\epsilon}$ & $Z_{\rm wind}$\\ \hline \hline
    \multirow{9}{*}{\makecell{High Z \\ ($\sim 10^{-2}$)}}
    & 1 ($Z$) & 0.02 & - & - & - &
    \multirow{9}{*}{\makecell{0.02}} \\ \cline{2-6}
    & 2 ($Z_{\mu}$) & 0.02 & 0.04 & 0.02 & 0.02  \\ \cline{2-6}
    & 3 ($Z_{\kappa}$)  & 0.02 & 0.02 & 0.04 & 0.02  \\ \cline{2-6}
    & 4 ($Z_{\epsilon}$) & 0.02 & 0.02 & 0.02 & 0.04 \\ \cline{2-6}
    & 5 ($Z_{\mu\kappa}$) & 0.02 & 0.04 & 0.04 & 0.02 \\ \cline{2-6}
    & 6 ($Z_{\mu\epsilon}$) & 0.02 & 0.04 & 0.02 & 0.04  \\ \cline{2-6}
    & 7 ($Z_{\kappa\epsilon}$) & 0.02 & 0.02 & 0.04 & 0.04  \\ \cline{2-6}
    & 8 ($Z_{\mu\kappa\epsilon}$) & 0.02 & 0.04 & 0.04 & 0.04  \\ \cline{2-6}
    & 9 ($2Z$)  & 0.04 & - & - & -  \\
     \hline \hline
     \multirow{9}{*}{\makecell{Low Z \\ ($\sim 10^{-3}$)}}
    & 10 ($Z$) & $10^{-3}$ & - & - & - &
    \multirow{9}{*}{\makecell{$10^{-3}$}} \\ \cline{2-6}
    & 11 ($Z_{\mu}$) & $10^{-3}$ & $2\times 10^{-3}$ & $10^{-3}$ & $10^{-3}$  \\ \cline{2-6}
    & 12 ($Z_{\kappa}$) & $10^{-3}$ & $10^{-3}$ & $2\times 10^{-3}$ & $10^{-3}$  \\ \cline{2-6}
    & 13 ($Z_{\epsilon}$) & $10^{-3}$ & $10^{-3}$ & $10^{-3}$ & $2\times 10^{-3}$  \\ \cline{2-6}
    & 14 ($Z_{\mu\kappa}$) & $10^{-3}$ & $2\times 10^{-3}$ & $2\times 10^{-3}$ & $10^{-3}$  \\ \cline{2-6}
    & 15 ($Z_{\mu\epsilon}$) & $10^{-3}$ & $2\times 10^{-3}$ & $10^{-3}$ & $2\times 10^{-3}$  \\ \cline{2-6}
    & 16 ($Z_{\kappa\epsilon}$) & $10^{-3}$ & $10^{-3}$ & $2\times 10^{-3}$ & $2\times 10^{-3}$  \\ \cline{2-6}
    & 17 ($Z_{\mu\kappa\epsilon}$) & $10^{-3}$ & $2\times 10^{-3}$ & $2\times 10^{-3}$ & $2\times 10^{-3}$  \\ \cline{2-6}
    & 18 ($2Z$) & $2\times 10^{-3}$ & - & - & - \\ \hline
\end{tabular}
\caption{Summary of the \texttt{MESA} models used in our study.  Columns give the different metallicity values used in each numerical experiment to dissect the effects of individual microphysics on stellar radii. See Sec.~\ref{sec:microphysics} for the notation.  The rows with empty entries are the fiducial models, for which all metallicity values are the same as in Column 1.} \label{tab:models}
\end{table*}

\subsection{Fiducial Models (Standard Microphysics)}
\label{sec:fiducial_model}

We first construct fiducial models, which employ standard
microphysics available in \texttt{MESA}, for two pairs of metallicity values:
$Z=(10^{-3}, 2\times 10^{-3})$ and $(0.02, 0.04)$ (Models 1, 9, 10, and
18 in Table~\ref{tab:models}, respectively).  Our motivation is to separately explore stellar radii around a characteristic `high'
metallicity value near solar ($Z \sim Z_{\odot} \simeq 0.02$) and a
characteristic `low' metallicity ($Z \sim 10^{-3}$), insofar as the
dominant physical processes at low- and high-metallicity may differ substantially.

For a $15M_{\odot}$ star, the EOS module in \texttt{MESA} uses the
OPAL tables (for $Z\leq0.04$; \citealt{Rogers+Nayfonov2002}) and HELM \citep{Timmes+swesty2000}.
 \texttt{MESA} uses tabulated opacities to construct stellar structures as functions of density and metallicity (\citealt{Lederer+09,Marigo+09}).  The radiative opacities are primarily from OPAL \citep{Iglesias+rogers1993,Iglesias+rogers1996}, with the low-temperature data from \citet{Ferguson+2005} and high-temperature,
Compton-scattering dominated regime from
\citet{Buchler+yueh1976}.  The electron conduction opacities are from
\citet{Cassisi+2007}.

The nuclear reaction rates used in \texttt{MESA} come from NACRE \citep{Angulo+1999} and JINA REACLIB
  \citep{Cyburt+2010}, plus additional tabulated weak reaction rates
  \citep{Fuller+1985,Oda+1994,Langanke+Martinez-Pinedo2000}. Screening is included via the
  prescription of \citet{Chugunov+2007}. The nuclear networks we use is
  \texttt{basic.net}, which includes the species $^{12}$C, $^{14}$N, $^{16}$O,
  $^{20}$Ne, $^{24}$Mg, and is extended to \texttt{co\_burn.net} and \texttt{approx21.net}, a 21-isotope network  in
  later evolutionary phases. This approach is sufficient to
    capture the bulk of the energy generation in the star until carbon
  depletion \citep[e.g.,][]{farmer:16}, but does not capture the details of
  nucleosynthesis.
  Nuclear neutrino loss rates
  are accounted for in the nuclear reaction rates; thermal neutrino
  loss rates are from \citet{Itoh+1996}.

  The initial metal isotope fractions are assumed to scale with the solar
abundances from \citet{Grevesse+Sauval1998}.  With the goal of
performing a ``controlled'' numerical experiment, we want to isolate
the effects of changing metallicity on stellar radii from those which
occur due to the associated changes in hydrogen and helium abundances.
We fix the hydrogen mass fraction $X=0.75$ in all of the models in
this paper, with the helium mass fraction then following from
$Y = 1-X-Z$.  This choice differs from the usual assumption of
changing both $X$ and $Y$ simultaneously with $Z$ (e.g.,
\citealt{Pols+98}, see also Sec.~\ref{sec:other_effects}).

  In addition to the microphysical processes whose effects are explored in this paper
(namely, photon opacities responsible for energy transport, equation of state and nuclear reaction rates), the metallicity in the stellar atmosphere impacts the mass-loss rate of the star $\dot{M}$ (most importantly, through radiation pressure on the atomic lines of iron-group elements; \citealt{Castor+75}).   It will prove convenient to define the ``wind metallicity'' $Z_{\rm wind}$ as that which enters $\dot{M}$.  All of our models employ the mass-loss prescription of
\citet{vink2001} and \citet{deJager+1988} for high and low temperature
phase, respectively \citep{Glebbeek+2009}, with metallicity-scaling $\dot{M} \propto Z_{\rm wind}^{0.85}$ \citep{Vink2005}.  We fix the value of $Z_{\rm wind}=0.02$ and $10^{-3}$ in all of the high-Z and low-Z models, respectively, rather than varying them directly with the overall metallicity of the model (see \citealt{Farmer+19} for a similar approach in a different context). We explore the quantitative impact of a varying wind metallicity as well in Sec~\ref{sec:other_effects}.

The treatment of convection also influences the stellar radius (e.g.,
\citealt{Dessart+2013}). We determine convectively unstable regions
using the Schwarzschild criterion and adopt mixing length theory
\citep{bohmvitense:58} with $\alpha_{\rm MLT}=2.0$
\citep{cox+giuli1968}. \citet{schootemeijer:19} showed that the radial
gradient of the (composition-dependent) mean molecular weight affects
the post-main sequence expansion (when using the Ledoux criterion) and
could lead to larger stellar radii. However, the hydrodynamics of
  convective boundary mixing is still an active topic of research
  \citep[e.g.,][]{anders:22a, anders:22b}, and which instability
  criterion is physically more appropriate when modeling stars should
  not depend on the metallicity-dependence of the microphysics we
  focus on here.
We adopt a step-function overshooting for the boundaries of any
convective region \citep{Paxton+2011}, with free parameters
(\texttt{f},\texttt{f\_0}) = (0.345,0.01) \citep{Brott+2011}.

\begin{figure}
    \centering
    \includegraphics[width=\columnwidth]{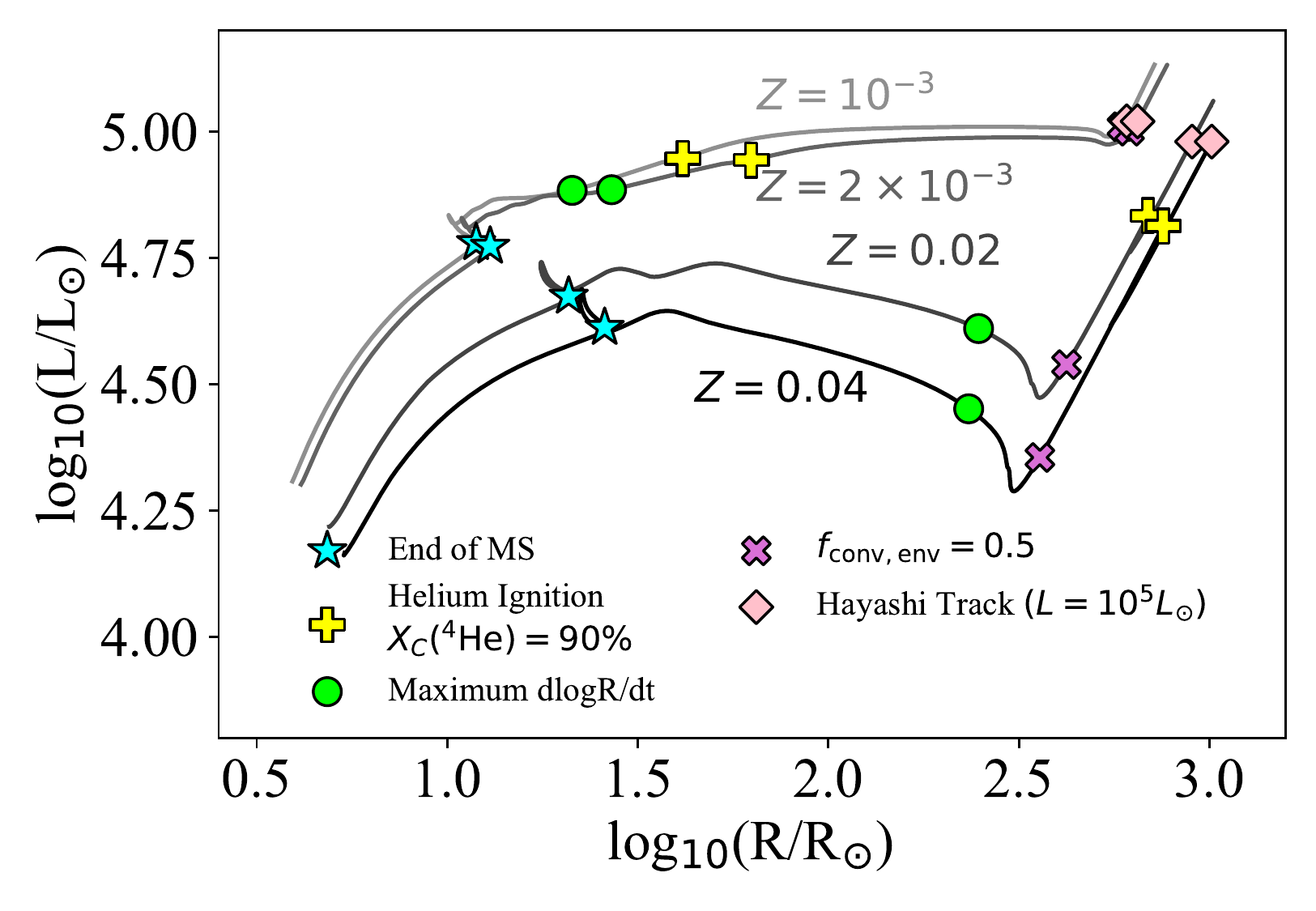}
    \caption{Tracks of stellar luminosity $L$ as a function of radius $R$ for our fiducial models with metallicity $Z=10^{-3}$
      , $Z=2\times 10^{-3}$, $Z=0.02$ and $Z=0.04$, from top to bottom. Several key epochs of interest are denoted with different symbols as marked.
}
    \label{fig:HRD_fiducial}
\end{figure}

Figure~\ref{fig:HRD_fiducial} shows evolutionary tracks of stellar luminosity $L$ versus stellar radius $R$ for our fiducial models, starting at the beginning of the MS.
Different symbols along the tracks denote key physical stages of interest discussed below.  Models with higher metallicity are less luminous at any given epoch (or, equivalently, possess larger radii at fixed luminosity).  This trend is well established in the literature \citep[e.g.][]{Garnett2002,Zahid+2011} but its origin in terms of the relative contributions of different microphysical input has to our knowledge not been elucidated in a systematic way.

\subsection{Models with Altered Microphysics Input} \label{sec:microphysics}

We focus on understanding what factors determine the radius differences between the
$Z = 10^{-3}$ and $2\times 10^{-3}$ models, and between the $Z = 0.02$
and $Z = 0.04$ models, respectively.  We do this by changing
separately the metallicity values which enter in
the opacity, EOS, and nuclear reactions in \texttt{MESA} with
customized routines (see, e.g., \citealt{walmswell:15} for
a related approach).
The metallicity values for each microphysical input are denoted in
Table \ref{tab:models} by the symbols $Z_{\kappa}$ for the metallicity
in the opacity calculations, $Z_{\mu}$ for the metallicity value in
the EOS calculations, and $Z_{\epsilon}$ for the metallicity in the
nuclear energy generation calculations, respectively. Models 2-9 and
models 11-17 cover all different permutations of the values of
$\{Z_{\mu}, Z_{\kappa}, Z_{\epsilon}\}$ away from the fiducial
high-metallicity models (1 and 9) and low-metallicity models (10 and 18),
respectively.

To implement the changes in $Z_{\kappa}$,
we implement a routine that mimics the default opacity treatment in
\texttt{MESA} with a separate input for metallicity.  Within this
routine, we rescale the total metallicity and the abundances of the
metal isotopes in our simulations
-- $^{12}$C, $^{14}$N, $^{16}$O,
$^{20}$Ne, $^{24}$Mg, which are used to interpolate opacities from the
tabulated values in \texttt{MESA}.  We choose the initial mass
fractions of each isotope coming from the higher metallicity fiducial model
in each category, which follows the solar abundances pattern, effectively setting $Z_{\kappa} = 2Z$. However, as
the star evolves, in the core the fraction of metals eventually
exceeds $2Z$ due to nuclear burning. Our models allow the
elemental abundances to change from their initial (possibly modified) values
self-consistently in later evolutionary stages, following the nuclear reaction networks in \texttt{MESA}.

Similarly, we create separate EOS routines that mimic the default
behavior but setting a floor $Z_{\mu} = 2Z$ in mean molecular
weight. In this case too, as the composition of the star evolves,
$Z_\mu$ is allowed to follow.  As described in the previous section, all of our models assume an initial hydrogen mass fraction $X = 0.75$ and helium mass fraction $Y = 1-X-Z$.

The nuclear reaction rates also depend on the stellar composition,
because they are proportional to (a power of) the density of isotopes
involved in each reaction.  To account for the change from $Z_{\epsilon} = Z$ to $Z_{\epsilon} = 2Z$, we double all the
reaction rates in the nuclear networks involving metal isotopes. These consist of 20
reactions out of the total 30 used in the largest (21-isotope) nuclear
network we adopt.  We keep the rate factors constant
throughout the entire stellar evolution, but these changes are most
relevant on the MS, when our simulation
  employ \texttt{basic.net}, where the
primordial metal isotopes enter into the CNO bicycle. After the main
sequence the star synthesizes its own metals, reducing the importance
of the primordial composition on the subsequent nuclear evolution.

Finally, we note that any aspects of the models unrelated to the
opacity, EOS, or nuclear reactions is treated in the same way as in
the fiducial models.

\section{Results} \label{sec:results}

In this section, we systematically analyze the stellar radius of each
model in Table~\ref{tab:models} at five notable epochs of stellar
evolution (highlighted by markers for the fiducial models of
Fig.~\ref{fig:HRD_fiducial}): 
{\it Terminal age main sequence} (Sec.~\ref{sec:MS}), corresponding to
the maximum radius during core hydrogen burning; {\it Hertzsprung Gap} (Sec:~\ref{sec:HG}), where the
rate of change of radius (dlog(R)/dt) is largest; {\it Helium core
  burning} (Sec.~\ref{sec:he_ignition}), where the central helium mass
fraction first reaches 90\%; {\it Beginning of
  Hayashi track} (Sec.~\ref{sec:hayashi_1}), where half of the
envelope mass is convective; and {\it Late Hayashi track}
(Sec.~\ref{sec:hayashi_2}), where the stellar luminosity reaches
$L=10^5L_{\odot}$.

\begin{figure*}
    \centering
    \includegraphics[width=2\columnwidth]{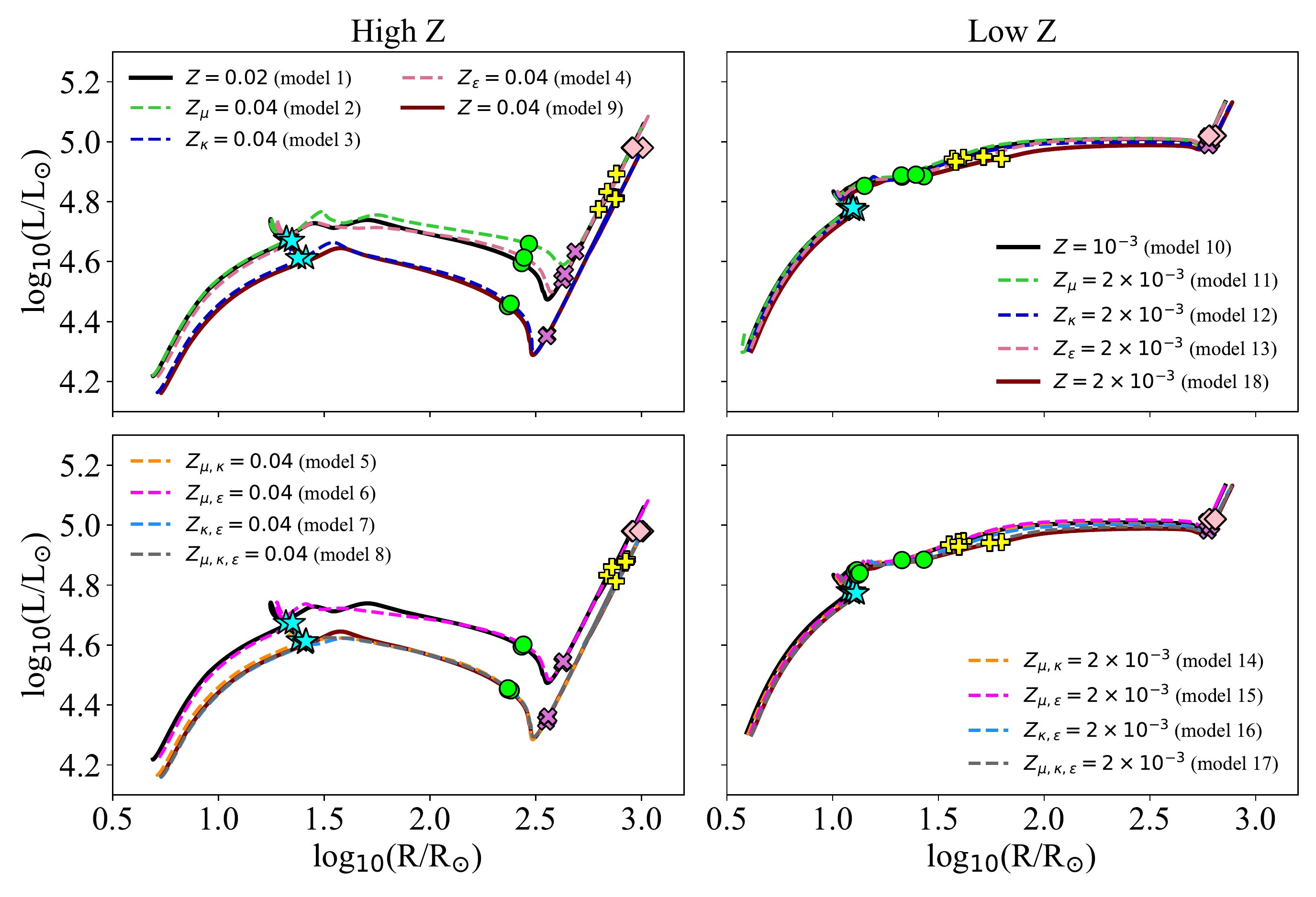}
    \caption{Luminosity-radius tracks for all models in
      Table~\ref{tab:models}, in both high metallicity
      $Z \sim 10^{-2}$ (left panels) and low metallicity
      $Z \sim 10^{-3}$ (right panels) models. The four fiducial models
      are superimposed with black and dark red solid lines, while the
      dashed lines show models where the metallicity for one (top
      panels) or more (bottom panels) microphysical input has been
      increased by a factor of 2 ($Z \rightarrow 2Z$). Key physical
      epochs are marked with the same symbols as in
      Fig.~\ref{fig:HRD_fiducial}.}
    \label{fig:HR_4panel}
\end{figure*}

Figure~\ref{fig:HR_4panel} shows the luminosity-radius evolution for each model in Table \ref{tab:models}. The left
panels show the `high' metallicity models ($Z \sim 10^{-2}$) while the
right panels show the `low' metallicity models ($Z \sim 10^{-3}$). The
top rows compare the fiducial models to those in which a single
microphysical metallicity value has been doubled (models \#1-4,
\#11-13) while the bottom row show cases in which two or more
microphysical metallicities have been doubled (models \#5-7, \#14-16).
We expect the lower metallicity tracks (black curves in each
panel; $Z=0.02$ and $10^{-3}$) to approach the
higher metallicity tracks (dark red curves; $Z=0.04$ and
$2\times 10^{-3}$) as we progressively increase each microphysical-metallicity $\{Z_{\kappa},Z_{\mu},Z_{\epsilon}\}$ from $Z$ to $2Z$.

In the high metallicity models ($Z \sim 10^{-2}$, left panels), opacity
has the largest impact on stellar radii throughout the MS
and on the Hayashi track: the $Z_{\kappa}, Z_{\mu\kappa}$ and
$Z_{\kappa\epsilon}$ models are closest to the $Z = 0.04$ fiducial
model. By comparison, doubling the metallicity in the EOS or nuclear
physics  has a less pronounced effect, producing stellar
tracks more similar to the fiducial $Z$ model. We also notice that the
$Z_{\mu\kappa\epsilon}$ model (for which all three microphysics
metallicities are set to the $2Z$ value), does not perfectly match the
fiducial $2Z$ model; we return to the origin of this discrepancy later in this section.

By contrast, in the low metallicity ($Z \sim 10^{-3}$) models, similar
trends are more challenging to discern from the figure alone because
the absolute difference in the radii between models are more modest,
due to the smaller absolute variation in metallicity
($\Delta Z=2\times10^{-3}$ vs. $\Delta Z=0.02$ in the high metallicity
case). Nevertheless, the offset of the symbols on the horizontal axis,
particularly in the `mixed' models shown in the lower right plot,
demonstrate the metallicity-sensitivity of the radii at key stages of
stellar evolution at low metallicity. The next section describes a
more quantitative method to dissect the metallicity dependencies of
the radius.

\subsection{Radius Matrix}
\label{sec:analysis}
Before proceeding to quantify our results for the partitioning of microphysical effects on stellar radii, we first introduce the notation used to present and interpret our findings. We define
\begin{equation}  \label{eq:matrix_2}
    \Delta_{2Z} \equiv \frac{R_{2Z}-R_Z}{R_Z} \ \ ,
\end{equation}
where $R_{Z}~(R_{2Z})$ is the radius of the fiducial model of metallicity
$Z~(2Z)$. Thus, $\Delta_{2Z}$ is the fractional change in the stellar
radius for the fiducial model of metallicity $2Z$ (models \# 9, 18) relative to the radius in the fiducial model of metallicity $Z$ (models \#1, 10) at each key epoch of stellar evolution (denoted by symbols in Fig.~\ref{fig:HRD_fiducial}).
Our aim is to quantitatively break down $\Delta_{2Z}$ into the
  contributions from different microphysics.

Similarly, we define
\begin{align}
    \Delta_{\{\kappa,\mu,\epsilon\}}
    \equiv\frac{R(Z,Z_{\kappa},Z_{\mu},Z_{\epsilon})-R_Z}{R_Z}  \ \ ,\label{eq:matrix_elem}
\end{align}
as the fractional change in the stellar radius
$R(Z,Z_{\kappa},Z_{\mu},Z_{\epsilon})$ of a given model with enhanced
metallicities $Z_{\kappa}$, $Z_{\mu}$, and/or $Z_{\epsilon} = 2Z$ (models
\#2-8 and \#11-17), again relative to the fiducial metallicity $Z$
(models \#1, \#10). For example, $\Delta_{\mu\kappa}$ is the result of
Eq.~\ref{eq:matrix_elem} for the model in which
$Z_{\mu} = Z_{\kappa} = 2Z$ but $Z_{\epsilon} = Z$, while
$\Delta_{\mu\kappa\epsilon}$ is the result of increasing all three
microphysics metallicities simultaneously
($Z_{\kappa} = Z_{\epsilon} = Z_{\mu} = 2Z$; models \#8, \#17).  We calculate
$\Delta$ values separately for the pairs of models at high ($Z\sim10^{-2}$) and
low metallicities ($Z\sim10^{-3}$).

We arrange our results in $3\times3$ matrices with entries $\Delta_{\{ \kappa,\mu,\epsilon \}}$ in percentage units, of the following form:
\begin{blockarray}{lcccc} \label{mat:format}
    & $\mu$ & $\kappa$ & $\epsilon$ & \\
\begin{block}{l[ccc]l}
  $\mu$ & $\Delta_{\mu}$ & $\Delta_{\mu\kappa}$ & $\Delta_{\mu\epsilon}$ & \ $\Delta_{\mu\kappa\epsilon}$ \\
  $\kappa$ & - & $\Delta_{\kappa}$ & $\Delta_{\kappa\epsilon}$ & \ Tr($\Delta) =\Delta_{\mu} + \Delta_{\kappa} + \Delta_{\epsilon} $\\
  $\epsilon$ & - & - & $\Delta_{\epsilon}$ & \ $\Delta_{2Z}$\\
\end{block}
\end{blockarray}

The matrix is symmetric by construction, and we report its trace
denoted as
Tr$(\Delta)= \Delta_{\mu} + \Delta_{\kappa}+\Delta_{\epsilon}$. We
generate one matrix for each of the key epochs of stellar evolution
shown with symbols in Fig.~\ref{fig:HRD_fiducial}-\ref{fig:HR_4panel}
(and subsequent figures), and discuss them
separately in the sections to follow.

Naively, one should expect $\Delta_{2Z}$ to equal
$\Delta_{\mu\kappa\epsilon}$, because activating all microphysics
inputs to their $2Z$ value should be equivalent to doubling the
metallicity of the fiducial model. However, as shown in
Fig.~\ref{fig:HR_4panel}, this expectation is not typically realized
in practice, as a result of numerical inaccuracies in our experiments.
Appendix \ref{appendix:numerical}  addresses two sources of such
errors: (1) uncertainties that arise from our non-standard numerical
implementation of metallicity-dependent microphysics in \texttt{MESA}
(as described in Sec.~\ref{sec:microphysics}); (2) uncertainties
associated with the adopted temporal or spatial grid resolution. As
the former source of error typically dominates over the latter, we
estimate the uncertainty $\delta$ on each matrix element $\Delta$ using the
radius error associated with our numerical implementation (entries of
the so-called ``null'' matrix; see Appendix \ref{appendix:numerical}). The latter source of error derives from the fact that the resolution requirements of every MESA model are dependent on the details of both physical and numerical inputs, \citep[e.g.,][]{Mehta2022}.  The routines used in the $Z_{ \mu,\kappa,\epsilon}$ and the $2Z$ models account for different input physics, so their resolution requirements can in principle differ (see further discussion in Appendix~\ref{sec:resolution}).
Also contributing to the difference between $\Delta_{2Z}$ and
$\Delta_{\mu\kappa\epsilon}$ is the different initial helium fraction
assumed in the $2Z$ versus $Z$ models (the helium fraction is fixed in
the models used to calculate $\Delta_{\mu\kappa\epsilon}$ at the
fiducial $Z$ model abundance but lower in the fiducial $2Z$ model;
Sec.~\ref{sec:other_effects}). All the  uncertainties we report below (entries inside parentheses) should not be considered as stochastic uncertainties, but rather systematic uncertainties.

One might also expect that $\Delta_{\mu\kappa\epsilon} =$ Tr($\Delta$), i.e. the fractional radius change that results from doubling all the microphysical
metallicities simultaneously would be equivalent to the sum of the changes that arise from activating them individually. This expectation
is also not satisfied in general. However, the origin of this discrepancy is at least partially for physical rather than numerical reasons.  Stellar evolution is a highly non-linear problem \citep[e.g.,][]{Kippenhahn&Weigert12}, so it
should not be surprising that non-linear effects can arise. For example, a different choice of opacity influences the stellar evolution by different
fractional amounts depending on the metallicity entering the EOS or
nuclear reaction rates.

\subsection{Main Sequence} \label{sec:MS}
Stars spend the majority of their lifetime on the MS,
corresponding to $\approx 1.5-1.9 \times 10^7$ years in our fiducial
$15\,M_\odot$ models, roughly independent of metallicity.
We first consider the impact of changing $\{Z_{\mu}$, $Z_{\kappa}$,
$Z_{\epsilon}\}$ on the stellar radius at the end of the MS. We define
the terminal age main sequence (TAMS) as a central hydrogen masfraction $X_{\rm c}(H) < 10^{-3}$, which correspond to the maximum
radial extent of stars before core hydrogen exhaustion
(cf.~Fig.~\ref{fig:HRD_fiducial} and \ref{fig:HR_4panel}) .

The radii matrices (Eq.~\ref{eq:matrix_elem}) for the high and low metallicity models are given, respectively, by:\\

\begin{center}
{\bf High $Z$}

\begin{blockarray}{ccccl} \label{mat:format}
   $\Delta$ (\%) & $\mu$ & $\kappa$ & $\epsilon$ & \\
\begin{block}{c[ccc]l}
  $\mu$ & 0.6(0.07) & 14.9(0.1) & 7.8(0.07) & \ \ $\Delta_{\mu\kappa\epsilon} = 24.5(0.1)$\\
  $\kappa$ & - & 14.1(0.07) & 23.2(0.07) & \ \ Tr($\Delta)=21.6(0.1)$ \\
  $\epsilon$ & - & - & 6.9(0.0) & \ \ $\Delta_{2Z}=24.3(0.0)$\\
\end{block}
\end{blockarray} \\

\newpage
{\bf Low $Z$}

\begin{blockarray}{ccccl} \label{mat:format}
   $\Delta$ (\%) & $\mu$ & $\kappa$ & $\epsilon$ &\\
\begin{block}{c[ccc]l}
  $\mu$ & 0.02(0.03) & 3.6(0.01) & 5.0(0.03) & \ \ $\Delta_{\mu\kappa\epsilon} = 8.9(0.01)$ \\
  $\kappa$ & - & 3.5(0.0) & 8.8(0.0) & \ \ Tr($\Delta)= 8.6(0.03)$ \\
  $\epsilon$ & - & - & 5.0(0.0) & \ \ $\Delta_{2Z}=8.8(0.0)$ \\
\end{block}
\end{blockarray} \\
\end{center}

The effect of increasing any individual ($\Delta_{\mu}$, $\Delta_{\kappa}$, $\Delta_{\epsilon}$) or combination ($\Delta_{\mu\kappa}$, $\Delta_{\kappa\epsilon}$, etc.) of
microphysical metallicities is to increase the TAMS radius, that is
all models exhibit a radius increase (positive $\Delta>0$). However,
not all the metallicity changes exert the same quantitative effect.
The $Z$-dependence of the opacity has the largest impact in the
high-$Z$ models ($\Delta_\kappa$ is the largest in
  the high $Z$ matrix), while nuclear reactions dominates in the low-$Z$
models, in agreement with Fig.~\ref{fig:HR_4panel} and previous
studies \citep[e.g.,][]{Farrell+21}. In both cases, the
metallicity-dependence of the EOS is subdominant.

We also find that the deviation from linearity when increasing $Z_{\mu}, Z_{\kappa},$ and $Z_{\epsilon}$
is relatively small: the sum of the radius increase due to each
  individual microphysics
is close to the radius variation obtained by changing all
microphysical-metallicity simultaneously. For example,
$\Delta_{\kappa} + \Delta_{\mu} \simeq \Delta_{\mu\kappa}$ in both
high- and low-metallicity models.

We now interpret each diagonal matrix element in terms of stellar
physics considerations. The opacity $\kappa$ controls the rate at
which photons carry energy through the stellar envelope to the
surface. The dominant source of opacity in the envelope
  are bound-free and bound-bound transition (e.g.,
  \citealt{Stothers&Chin93}). These increase with metallicity, thus a
  higher metallicity $Z_\kappa$ (higher envelope opacity) results in a
  lower stellar luminosity (e.g., \citealt{Kippenhahn&Weigert12}). For
  a fixed nuclear burning rate and EOS, a lower stellar luminosity
  therefore requires a lower core temperature, $T_c$. Therefore, from
  the virial theorem, $kT_c \propto GM\mu m_p/R$, a higher
  $Z_{\kappa}$ will result in a larger radius, consistent with
  $\Delta_\kappa=+14\%$ in the high-Z models.
By contrast, in low-metallicity stars, electron scattering opacity,
$\kappa\approx 0.2(1+X)$ cm$^2$ g$^{-1}$, plays a larger relative role.  Hence, we would not expect as significant of a radius change in from changing $Z_{\kappa}$ in our $Z \sim 10^{-3}$ scenario, consistent with the smaller value of $\Delta_\kappa=+0.8\%$.

Next, consider the effects of the nuclear burning metallicity,
$Z_{\epsilon}$.  In massive stars, the CNO cycle (e.g.,
\citealt{Bethe39}) dominates hydrogen burning in the core, with a
specific nuclear burning rate
$\epsilon_{\rm CNO}\propto Z_{\epsilon}T^{\alpha}$, where
$\alpha\approx 20$.  The large value of the exponent $\alpha$
concentrates the burning in the central region, so the temperature
  entering $\epsilon_{\rm CNO}$ 
can be approximated as the central value $T_c$.
At a given point on the MS, the envelope opacity and hence the stellar luminosity $L$ is approximately fixed. Assuming the star is in thermal equilibrium, i.e. $L \simeq L_\mathrm{nuc} \propto \epsilon_{\rm CNO}(T_{\rm c})$, where $L_\mathrm{nuc}$ is the luminosity from nuclear burning, which is proportional to the energy generation rate from the
CNO cycle (times the amount of fuel available at that point on the MS).  Taking $L = \rm{constant}$ and hence
$\epsilon_{\rm CNO} = {\rm constant}$ then implies
$T_{\rm c} \propto Z_{\epsilon}^{-1/\alpha}$ and hence from the virial
theorem $R \propto T_c^{-1} \propto Z_{\epsilon}^{1/\alpha}$.  Thus,
doubling $Z_{\epsilon}$, should act to increase the radius by a factor
$\sim 2^{1/20}-1 \sim 4\%$.  This roughly agrees with our numerical results, $\Delta_{\epsilon} \approx 7$\% and $5$\% in the high- and low-$Z$ models, respectively.

Finally, consider the effect of the EOS-metallicity.  Gas pressure dominates in our stellar models: the mass averaged gas to total pressure ratios on the MS are $\langle\beta\rangle=P_{\rm gas}/P_{\rm tot}=0.81$ and $0.85$ for $Z=10^{-3}$ and $Z=0.02$ fiducial models, respectively.   Metallicity enters the EOS primarily through the dependence $P_{\rm gas} \propto \rho T/\mu$ on the mean molecular weight, $\mu$.  For fully ionized gas, the latter can be written
\begin{equation} \label{eq:mu}
    \mu \simeq \frac{1}{2X + \frac{3}{4}Y + \frac{1}{2}Z_{\mu}}
    = \frac{1}{\frac{3}{4} - \frac{5}{4}X - \frac{1}{4}Z_{\mu}},
\end{equation}
where in the final equality we have used $X+Y+Z_{\mu} = 1$ to express everything in terms of the hydrogen mass fraction $X$, which is fixed in our models.  An increase in the EOS-metallicity $Z_{\mu}$ by an
amount $\delta Z_{\mu}$ thus increases the value of $\mu$, such that
for small changes $\delta \mu/\mu \sim -(1/4)\mu \delta Z_\mu = 0.15\delta Z_\mu$
(where we take $\mu \simeq \mu_\odot \simeq 0.6$). From the
virial theorem $R \propto \mu/T_{\rm c}$, such that for the approximately fixed central temperature $T_{\rm c}$ set by nuclear energy generation
(see above), we expect $R$ to increase with increasing $\mu$, as $\delta R/R \sim 0.15 \delta Z_{\mu}$.  These expected changes are roughly borne out by our numerical results,
$\Delta_{\mu} \approx 0.6$\% and $0.02$\%, for
$\delta Z_{\mu} = 0.02$ and $\delta Z_{\mu} = 10^{-3}$, respectively, to within a factor of $\sim 2$.

\subsection{Hertzsprung Gap} \label{sec:HG}
\begin{figure}
    \centering
    \includegraphics[width=\columnwidth]{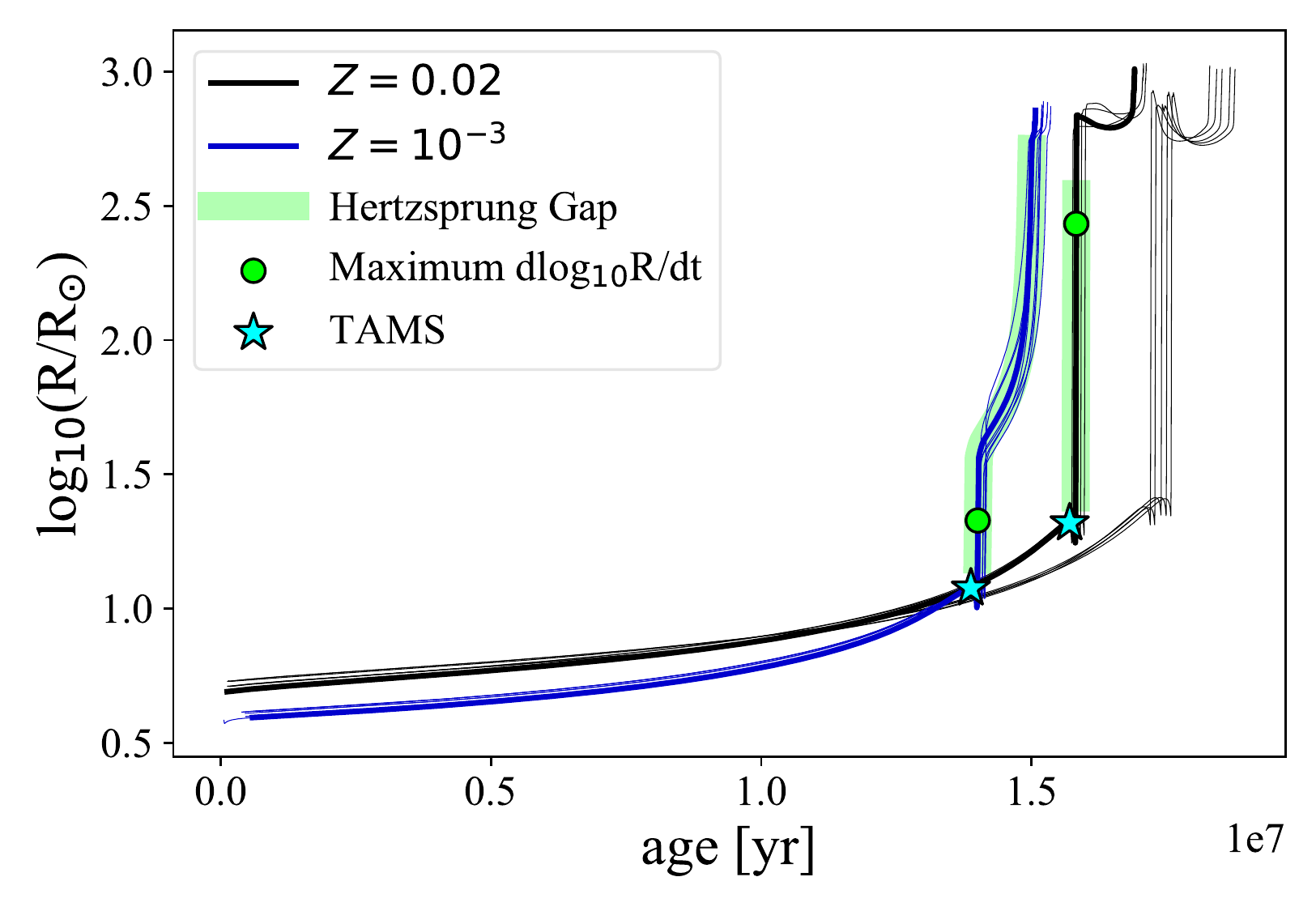}
    \caption{Stellar radii as a function of age for the high-Z (thick
      black line) and low-Z (thick blue line) fiducial models,
      demonstrating the rapid evolution during the post-MS evolution.
      The Hertzsprung Gap phase is shaded green with circles denoting
      the point of fastest radial expansion (when $d\log_{10}R/dt$ is
      maximum). Blue star symbols denote the largest radii achieved on the
      MS.  Thin blue and black lines show for comparison the
      non-fiducial low-Z and high-Z models, respectively, from
      Fig.~\ref{fig:HR_4panel}.
      }
    \label{fig:t_r_HG}
\end{figure}

Most interacting massive binaries experience mass transfer when the donor star crosses the ``Hertzsprung Gap'' (HG), as this is the phase of largest radial expansion (\citealt{vandenheuvel:69, sana:12, renzo:19walk}); see Fig.~\ref{fig:t_r_HG} for the evolution of stellar radius with age for our fiducial models.  The time a star takes to cross the Hertzsprung Gap is therefore important in determining the timescale for these mass
transfer episodes (although ultimately the dynamical stability of mass
transfer is determined by the reactions of the stellar radii
and the orbital separation to the changes in the stellar masses, not radii; e.g., \citealt{soberman:97, vigna-gomez:20}).  If the donor is intrinsically evolving ``fast'' (e.g., on a
thermal timescale), this forces an at least equally ``fast'' mass
transfer timescale in a binary. Conversely, if the donor star is
evolving ``slowly'' (e.g., on a nuclear timescale, \citealt{klencki:21b}), then other process
(e.g., the structural reaction of the secondary or the secular
evolution of the orbit) can determine the binary mass-transfer timescale.

These reasons motivate us to explore the metallicity effects on stellar radii during the epoch of fastest radial expansion, i.e. when d($\log_{10}$R)/dt is maximum (green dots in
Figs.~\ref{fig:HR_4panel}, \ref{fig:t_r_HG}).  As illustrated by the matrices below, doubling each microphysics metallicity value acts to reduce the
stellar radius at this stage (negative $\Delta$ values), except for
the EOS-metallicity at high-Z and nuclear-metallicity at low-Z.  Opacity plays the largest role in changing the radius,
with $\Delta_{\mu} \sim$11\% and $\sim$34\% in the high-Z and low-Z
models, respectively.  Unfortunately, however, the uncertainties in
the low-Z model results are similar in magnitude to the matrix entries (see Appendix~\ref{appendix:numerical}), so no firm conclusions can be drawn in this case.

\begin{center}
{\bf High $Z$}
\begin{blockarray}{ccccl} \label{mat:format}
   $\Delta$ (\%) & $\mu$ & $\kappa$ & $\epsilon$ & \\
\begin{block}{c[ccc]l}
  $\mu$ & 8.1(9.9) & -11.2(4.1) & 1.8(9.9) & \ \ $\Delta_{\mu\kappa\epsilon} = -14(4.1)$\\
  $\kappa$ & - & -11.4(-1.2) & -11.4(-1.2) & \ \ Tr($\Delta)= -0.8(8.7)$ \\
  $\epsilon$ & - & - & 2.5(0.0) & \ \ $\Delta_{2Z}=-14(-3.4)$\\
\end{block}
\end{blockarray} \\
{\bf Low $Z$}
\begin{blockarray}{ccccl} \label{mat:format}
   $\Delta$ (\%) & $\mu$ & $\kappa$ & $\epsilon$ & \\
\begin{block}{c[ccc]l}
  $\mu$ & -0.7(6.4) & 40(-42) & -39(6.4) & \ \ $\Delta_{\mu\kappa\epsilon} = -37(-42)$\\
  $\kappa$ & - & -34(-42) & -38(-42) & \ \ Tr($\Delta)= -18(-35)$ \\
  $\epsilon$ & - & - & 16.3(0.0) & \ \ $\Delta_{2Z}=27(-1.7)$\\
\end{block}
\end{blockarray} \\
\end{center}

\subsection{Helium core burning} \label{sec:he_ignition}
After a brief phase of contraction at H exhaustion (causing the
  so-called ``Henyey hook''), massive stars rapidly ignite core He
burning; this is the last long-lived phase of evolution, about
$\sim{}10\%$ of the total lifetime. In an observed population of stars, this
is typically the most evolved phase for which a statistically
significant number of stars can be assembled. Our $15\,M_\odot$ models
take about $10^5$\, yrs to evolve from a central $^4\mathrm{He}$ mass
fraction of $X_c(^4\mathrm{He})\sim{}1$ to $\sim{}0.9$, and $10^6$\,
yrs to reach He-depletion ($X_c(^4\mathrm{He})=0.01$ in the center).

Whether a star in this phase appears as a blue supergiant (BSG) with a
small radius or a red supergiant (RSG) with a large radius is not
fully understood. The answer depends sensitively on the previous
evolution \citep[e.g.,][and references
therein]{woosley:88,Kippenhahn&Weigert12, Farrell+21}, specifically
the assumptions made to model convective boundary mixing and
semiconvection \citep[e.g.,][]{langer:89, Brott+2011,
  schootemeijer:19}, wind mass loss and rotation
\citep[e.g.,][]{renzo:17,farrell:21,Sabhahit+2021}. Observations of
massive stars in the Galaxy, Large and Small Magellanic clouds suggest
a metallicity dependence of the BSG/RSG ratio \citep{klencki:2020a}.

\begin{figure}
    \centering
    \includegraphics[width=\columnwidth]{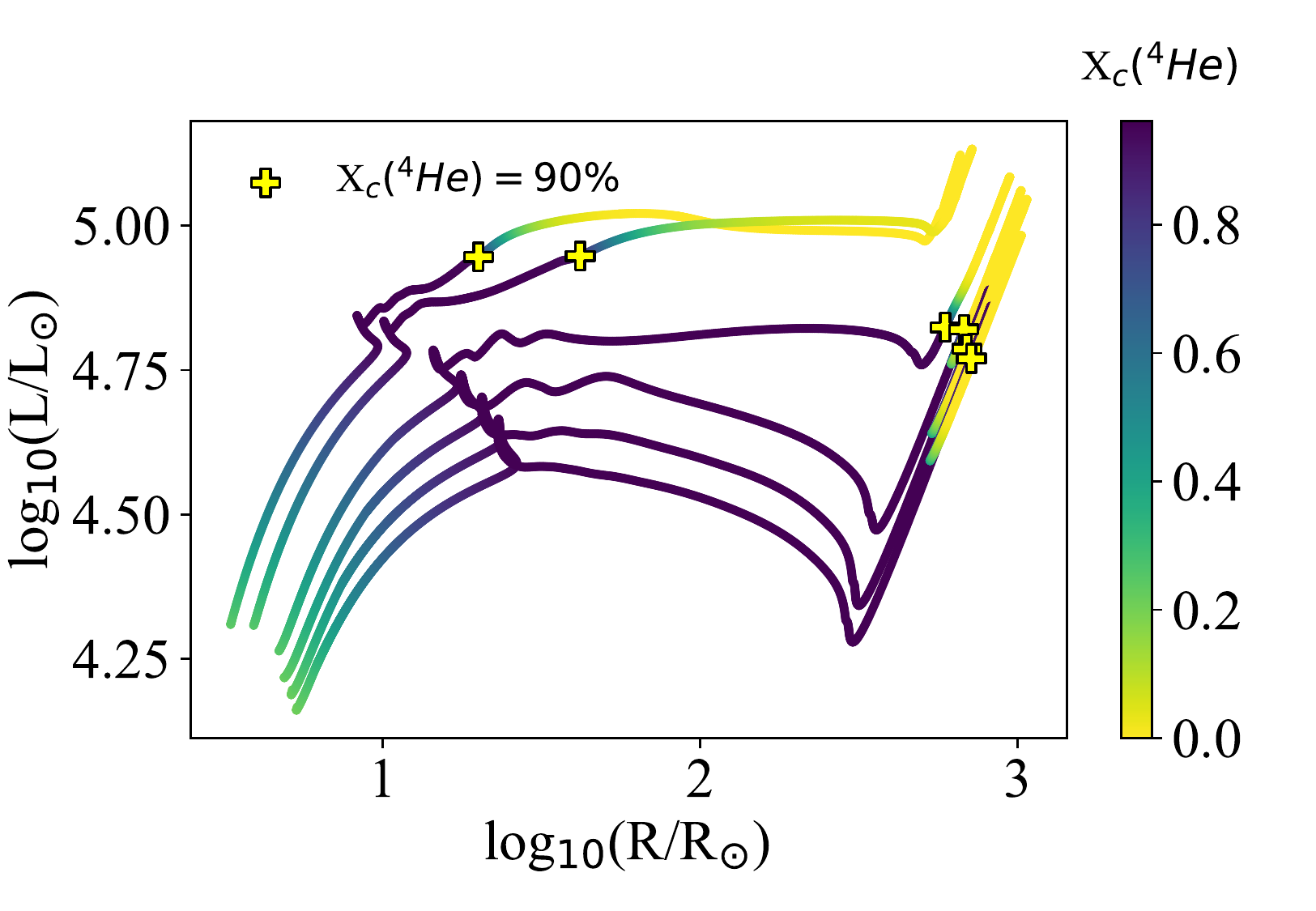}
    \caption{Luminosity-radius tracks showing the helium core burning
      phase for a range of metallicities --
      $Z=10^{-4}, 10^{-3}, 0.01, 0.02, 0.03$ and $0.04$ from top left
      to bottom right, calculated under the fiducial model
      assumptions. The color of each track represents the mass
      fraction of ${^4}$He in the core. Gray points represent the
      radius of helium ignition, defined as
      X$_{\rm c}({^4\rm He})=90\%$.
        }
    \label{fig:HRD_He_ign}
\end{figure}

To place this phase into context, Figure~\ref{fig:HRD_He_ign} shows $L$-$R$ tracks for several models calculated under fiducial model assumptions (Sec.~\ref{sec:fiducial_model}), for a wide range of metallicities, $Z=10^{-4}, 10^{-3}, 0.01, 0.02, 0.03$, $0.04$.  We color the tracks according to the core $^4$He mass fraction. He ignition occurs with a drastic decrease in $X_c(^4\mathrm{He})$, i.e. the quickest transition in colors in Fig.~\ref{fig:HRD_He_ign}.  Our 15\,$M_\odot$ low metallicity stars ($Z=10^{-4}, 10^{-3}$) burn He early in the Herzsprung Gap, corresponding to a high effective temperature of a BSG, while in the higher metallicity stars, He ignition occurs at a lower effective temperature during the RSG phase.

The matrices below compare stellar radii at the epoch when $X_c(^4\mathrm{He}) = 0.9$ (red diamonds in Fig.~\ref{fig:HRD_He_ign}); however, because the radius changes by $\lesssim 5\%$ throughout the He core burning phase, our results are not sensitive to this precise definition.

\begin{center}
{\bf High $Z$}

\begin{blockarray}{ccccl} \label{mat:format}
   $\Delta$ (\%) & $\mu$ & $\kappa$ & $\epsilon$ & \\
\begin{block}{c[ccc]l}
  $\mu$ & -7.4(-6.4) & 6.5(-3.9) & 7.2(-6.4) & \ \ $\Delta_{\mu\kappa\epsilon} = -0.2(-3.9)$\\
  $\kappa$ & - & 6.6(-8.9) & -2.9(-8.9) & \ \ Tr($\Delta)= 11.7(-15.3)$ \\
  $\epsilon$ & - & - & 12.6(0.0) & \ \ $\Delta_{2Z}=4.7(0.9)$\\
\end{block}
\end{blockarray} \\

{\bf Low $Z$}

\begin{blockarray}{ccccl} \label{mat:format}
   $\Delta$ (\%) & $\mu$ & $\kappa$ & $\epsilon$ & \\
\begin{block}{c[ccc]l}
  $\mu$ & -10(13) & -14(-21) & -4.7(13.4) & \ \ $\Delta_{\mu\kappa\epsilon} = 33(-21)$\\
  $\kappa$ & - & -7.1(-25) & -4.0(-25) & \ \ Tr($\Delta)=8.3(-11.2)$ \\
  $\epsilon$ & - & - & 25.6(0.0) & \ \ $\Delta_{2Z}=51(-5.7)$\\
\end{block}
\end{blockarray} \\
\end{center}

We see that the nuclear-metallicity has the highest impact on stellar radius in both the high-Z and low-Z cases, at
$\Delta_{\epsilon} \approx 18\%$ and $\Delta_{\epsilon}\approx 26\%$,
respectively, followed by EOS-metallicity and then the opacity.  Unfortunately the uncertainties on the matrix entries are comparable to or larger than their physical values, particularly in the low-Z models (see Appendix \ref{sec:resolution}).

The large differences between $\Delta_{\mu\kappa\epsilon}$ and Tr($\Delta$) in both the high- and low-Z models implicate the presence of non-linear interactions between the impact of the three microphysical effects.  There is also a smaller but still significant difference between $\Delta_{\mu\kappa\epsilon}$ and $\Delta_{2Z}$, which is noticeably larger in the high-Z case. This discrepency may be related to the difference in the initial helium mass fraction of the $2Z$ versus
$Z_{\kappa} = Z_{\mu} = Z_{\epsilon} = 2Z$ models (Sec.~\ref{sec:other_effects}).

\subsection{Beginning of Hayashi Track (Half Convective Envelope)} \label{sec:hayashi_1}

A key element determining the stability of mass transfer -- and the
  outcome of common envelope events -- in a binary system is the density stratification of the donor star's
envelope. Efficient convection in RSG envelopes enforces a
flat entropy profile: if mass is removed from the surface, the
envelope responds expanding, leading more likely to unstable mass
transfer (although thin radiative surface layers can greatly stabilize
the mass transfer, \citealt{pavlovskii:17, marchant:21}).

As stars evolve redward on the HR diagram, they develop a convective
envelope which grows inward. Therefore, the fraction of envelope mass
that is convective ($f_{\rm conv, env}$) grows at this stage.  Below, we compare the model stellar radii at the point when the convective mass of the envelope has grown to half of the total envelope mass, or $f_{\rm conv, env}$=0.5 (purple
crosses in Figure~\ref{fig:HR_4panel}). We obtain
the envelope mass by subtracting the mass of the He core from the total
mass of the star, where the helium core boundary is defined at the
outermost location where the hydrogen mass fraction is $\leq$ 0.01
and the helium mass fraction is $\geq 0.1$.
The $f_{\rm conv, env}$=0.5 epoch occurs before the He core burning phase (Sec.~\ref{sec:he_ignition}) in the high-Z models and after in the low-Z models.

\begin{center}
{\bf High $Z$}

\begin{blockarray}{ccccl} \label{mat:format}
   $\Delta$ (\%) & $\mu$ & $\kappa$ & $\epsilon$ & \\
\begin{block}{c[ccc]l}
  $\mu$ & 0.2(3.9) & -1.8(1.3) & 1.9(3.9) & \ \ $\Delta_{\mu\kappa\epsilon} = -1.8(1.3)$\\
  $\kappa$ & - & -7.9(-0.3) & -2.1(-0.3) & \ \ Tr($\Delta)= -11.7(3.6)$ \\
  $\epsilon$ & - & - & -4.0(0.0) & \ \ $\Delta_{2Z}=-8.3(1.0)$\\
\end{block}
\end{blockarray} \\

{\bf Low $Z$}

\begin{blockarray}{ccccl} \label{mat:format}
   $\Delta$ (\%) & $\mu$ & $\kappa$ & $\epsilon$ & \\
\begin{block}{c[ccc]l}
  $\mu$ & -0.3(0.4) & -0.2(-0.7) & 0.03(0.4) & \ \ $\Delta_{\mu\kappa\epsilon} = 1.1(-0.7)$\\
  $\kappa$ & - & 0.06(-1.2) & 0.3(-1.2) & \ \ Tr($\Delta)= 0.3(-0.8)$ \\
  $\epsilon$ & - & - & 0.6(0.0) & \ \ $\Delta_{2Z}=1.1(-0.1)$\\
\end{block}
\end{blockarray} \\
\end{center}

We see that the opacity-metallicity has the largest effect on radii at this phase in the high-Z models, nuclear-metallicity is a close
second, and the EOS-metallicity is the least important, within
reasonable degrees of uncertainties.  However the results in the low-Z case are again obscured by large numerical uncertainties (Appendix \ref{sec:appendix_a}).

\subsection{Late Hayashi Track} \label{sec:hayashi_2}
Neglecting possible late, dynamical mass loss events
\citep[e.g.,][]{quataert:12, shiode:14, khazov:16, fuller:17},
our $15\,M_\odot$ models
spend their final evolutionary stages on the Hayashi track, at temperatures
$3200\,\mathrm{K}\lesssim T_{\rm eff} \lesssim 4000\,\mathrm{K}$.
Since the envelope structure is essentially frozen during the final
stages of stellar evolution due to high neutrino luminosity of the core \citep{fraley:68}
, its radius will be similar to that at the time of core collapse and (potential) supernova explosion (however, see \citealt{quataert:12}).
The final stellar radius has important implications for the early-time
light curves of supernovae (e.g., \citealt{Nakar2010,Piro2013,morozova:18,goldberg:20,gill:22}).

Comparing our models on the Hayashi track at the point they reach a
common luminosity, $L=10^5L_{\odot}$, we obtain the following
matrices:

\begin{center}
{\bf High $Z$}
\begin{blockarray}{ccccl} \label{mat:format}
   $\Delta$ (\%) & $\mu$ & $\kappa$ & $\epsilon$ & \\
\begin{block}{c[ccc]l}
  $\mu$ & -0.1(0.03) & 12.7(0.3) & 1.2(0.03) & \ \ $\Delta_{\mu\kappa\epsilon} = 9.2(0.3)$\\
  $\kappa$ & - & 12.5(-0.5) & 12.3(-0.5) & \ \  Tr($\Delta)=13.3(-0.5)$ \\
  $\epsilon$ & - & - & 0.9(0.0) & \ \ $\Delta_{2Z}=12.6(0.0)$ \\
\end{block}
\end{blockarray}
\\

{\bf Low $Z$}

\begin{blockarray}{ccccl} \label{mat:format}
  $\Delta$ (\%) & $\mu$ & $\kappa$ & $\epsilon$ &\\
\begin{block}{c[ccc]l}
  $\mu$ & -0.11(-0.06) & 5.7(-0.8) & -0.6(-0.06) & \ \ $\Delta_{\mu\kappa\epsilon} = 6.3(-0.8)$ \\
  $\kappa$ & - & 6.2(-0.9) & 5.8(-0.9) & \ \ Tr($\Delta)=5.2(-1.0)$ \\
  $\epsilon$ & - & - & -0.8(0.0) & \ \ $\Delta_{2Z}=6.8(0.0)$ \\
\end{block}
\end{blockarray}  \\
\end{center}
In both high-$Z$ and low-$Z$ models, the opacity-metallicity has the biggest effect on the stellar
radius.  To quantitatively understand this behavior, we can approximate the star as being almost fully convective. Since convection in the bulk of RSG envelopes is efficient, the entropy profile will be constant and
$P \propto \rho^{5/3}$. We further assume an ideal gas EOS,
$P=P_{\rm gas} = \rho T/\mu m_p$, following \citep{Hayashi61}. These approximations are well-justified for our models: the fraction by radius of the envelope which is still radiative at $L = 10^5L_{\odot}$ is less than 0.1\% in the high $Z$ models and $\lsim 8\%$ for the low$-Z$ models.  Even at this late
stage of stellar evolution, gas pressure dominates over radiation
pressure in our 15\,$M_\odot$ models
($\langle\beta\rangle \equiv P_{\rm gas}/P_{\rm tot}\gsim0.7$). The
central pressure and temperature for a $\gamma = 5/3$ polytrope are
given by, $P_{\rm c} \simeq 0.77 GM^{2}/R^{4}$ and
$kT_{\rm c} \simeq 0.54 GM\langle\mu\rangle m_p/R$, where $\langle\mu\rangle$ is the
average mean molecular weight.

The photosphere occurs roughly where the average photon mean-free path $1/\kappa\rho$ equals the atmospheric scale-height, $H \equiv c_{\rm s,ph}^{2}/g$, where $g = GM/R^{2}$ is the surface gravity and $c_{\rm s,ph}$ is the sound speed at the photosphere. The dominant opacity source near the photosphere is the bound-free absorption by $H^{-}$ ions, which we approximate as (e.g., \citealt{Kippenhahn&Weigert12})
\begin{equation}
    \kappa_{\rm H^{-}} \simeq 2.5\times 10^{-31}\left(\frac{Z_{\kappa}}{Z_{\odot}}\right)\rho_{\rm ph}^{1/2}T_{\rm eff}^{9}\,{\rm cm^{2}\,g^{-1}},
\end{equation}
where the linear dependence with $Z_{\kappa}$ assumes that the electrons at these low temperatures are from singly-ionized alkali metals rather than hydrogen or helium.
Combining the above relations, we obtain
\begin{equation}
    R \propto L^{49/102}M^{-14/51}\langle\mu\rangle^{-26/51}Z_{\kappa}^{8/51}.
    \label{eq:R_ph}
\end{equation}
Equation \ref{eq:R_ph} predicts that doubling $Z_{\kappa}$ at fixed
$\{M,R,\langle\mu\rangle\}$ should increase $R$ by $\approx 11\%$,
roughly consistent with the high$-Z$ matrix entries above, e.g.
$\Delta_{\kappa} \approx 12.5\%$.  The agreement for the low$-Z$
models is not as good (though still within a factor $\lesssim 2$),
possibly due to the star not being fully-convective or a break-down of
the assumption that alkali metals supply the electrons which contribute to the $H^{-}$ opacity.

The much weaker dependence of stellar radius on $Z_{\mu}$ can be understood because the mean molecular weight is dominated by hydrogen and helium in the envelope.  For a fully ionized stellar interior one would predict $\delta \mu/\mu \sim 0.9\delta Z_{\mu}$ (Eq.~\ref{eq:mu} and surrounding discussion), such that from Eq.~\ref{eq:R_ph} one predicts $\delta R/R \sim -(26/51)(0.9 \delta Z_{\mu}) \sim -0.5\delta Z_{\mu};$ this is consistent with sign, but not the magnitude, of the $\Delta_{\mu}$ entry for the high-$Z$ models.  However, the stellar envelope will not be completely ionized given the lower temperature of the giant star envelope and hence a more detailed consideration of ionization-state-dependence of $\langle\mu\rangle$ is needed to make a quantitative prediction.  Finally, the weak dependence on $Z_{\epsilon}$ we observe is also expected, given that the properties of the nuclear energy source do not enter to first order in setting the radius of a fully convective star.

\section{Discussion} \label{sec:discussion}

The previous sections have quantified the relative importance of
different microphysical processes on the metallicity-dependence of
massive star radii.  This section describes some implications,
applications, and caveats of our results.

\subsection{Microphysics Error Propagation}
\label{sec:error}

One application of the general technique developed in this paper is as
a tool to propagate theoretical or experimental uncertainties in the
microphysics inputs
into corresponding
theoretical uncertainties in modelled stellar radii. We briefly discuss the
implications of our findings in this regard for each microphysics
input.
\paragraph*{Opacity.}  We have found that the metallicity-dependence
of the opacity has the largest impact on stellar radii during most
evolutionary epochs.  This implies that theoretical estimates of
stellar radii are particularly sensitive to uncertainties in opacity.
Our stellar models employ the OPAL opacity tables (see Sec.~\ref{sec:method}
for details); however, the knowledge of atomic physics in stellar environment is still evolving. As an example of the potential magnitude of such changes, the updated OPAL tables in \citet{Iglesias+rogers1996} yielded up to a 20\% increase in opacity versus the previous baseline.  Furthermore, the wavelength-dependent opacity of iron was recently measured by \cite{Bailey+2015} to be $\sim$75\% higher than in the
  \texttt{OP} and \texttt{OPAL} tables for solar conditions; this has a significant impact on the location of the $\beta$ Cep pulsational
  instability strip on the Hertzsprung-Russell diagram
  \citep{Moravveji2016}. If more generally representative of opacity
  uncertainties, this would imply the theoretical uncertainty on
  stellar radii of a MS star is comparable to that expected in moving from
  Milky-Way ($Z \sim Z_{\odot}$) to SMC-like ($Z \sim 0.1Z_{\odot}$)
  metallicities.

  \paragraph*{Equation of State.} Unlike opacity, uncertainties in the OPAL equation of state used in our numerical experiments in the regimes of temperature and density relevant to the evolution of a 15M$_{\odot}$ star are typically small ($\sim 10\%$; \citealt{Rogers+Nayfonov2002}).  Furthermore, our experiments indicate that the EOS-metallicity has the weakest impact on stellar radii.  We therefore conclude that the EOS is not a major contributor to the uncertainty in massive star radii.

  \paragraph*{Nuclear Reactions.} Our results show that in the early evolutionary stages of low-Z stars, nuclear reactions dominate the metallicity-dependence of the stellar radius.  However, the experimental uncertainties in the nuclear reactions rates at a given temperature and density in \texttt{MESA} are typically $\sim 10\%$ \citep{Sallaska+2013,Fields+2018} for the CNO reactions (though we note the bottleneck reaction $^{14}$N(p$,\gamma$)$^{15}$O of the cycle has undergone rate changes as large as a factor of $\sim 2$; e.g., \citealt{Formicola+03,Imbriani+2004}).  However, because the metallicity-sensitivity of stellar radii at low-metallicity is low in an absolute sense, nuclear reaction uncertainties should not be a major source of error.

\subsection{Observational Implications}

The Introduction enumerated several applications which depend on stellar radii and being able to quantify their accuracy.  Here, we expand this discussion guided by the results of our numerical experiments.

\paragraph*{Timing of Shock Breakout.} The $15\,M_\odot$ single non-rotating
stellar progenitors explored in our models reach core-collapse as RSG
with large radii
(cf.~Fig.~\ref{fig:HRD_fiducial}-\ref{fig:HRD_He_ign}).  Assuming such stars undergo successful explosions, they are likely to produce Type IIP supernovae (SNe).  Core collapse SNe are sources of thermal neutrino emission (e.g., \citealt{Burrows&Lattimer86}) and of gravitational waves (GW; e.g., \citealt{ott:09}), within seconds of core bounce.  The sensitivity of GW searches is sensitive to the time-window of the data-stream being searched, particularly for a weak signal.  Absent a coincident neutrino detection, electromagnetic observations of the SN shock breaking out of the stellar surface can provide an estimate of the core collapse time (e.g., \citealt{goldberg:20,gill:22}); however, the accuracy of the obtained time delay relies on knowledge of the stellar radius.

\paragraph*{Supernova Light Curves.} The initial rise-time and
luminosity of core collapse SNe light curves depends sensitively on
the stellar radius at the time of explosion and any pre-explosion mass
loss \citep[e.g.,][]{Nakar2010,Piro2013,Valenti+2016,Morozova2017,morozova:18}. Our
results show that during the final evolutionary stages, the radius
uncertainties are dominated by $Z$-dependent opacity, being $\sim$6\%
and 12\% in subsolar and solar case, respectively. Other microphysics
have $<1$\% effects in stellar radii, negligible compared to the
uncertainties in progenitor evolution \citep[e.g.,][]{OConnor2011, Sukhbold2018, Laplace+2021, Renzo2021} and
macrophysical effects \citep[e.g.,][]{Woosley2002, Langer2012,
  renzo:17, Davies2018, zapartas+2021, Zapartas2021b}.  However, the microphysics effects explored in this paper may be swamped by the three-dimensional effects not captured by a 1D stellar evolution model (e.g., \citealt{Goldberg+22}).

\paragraph*{Other Applications.}
Although this work focuses on massive stars, our general technique
could also be applied to assess uncertainties on lower-mass stellar
radii. They are relevant in the context of inferring stellar
properties through eclipsing star
\citep{MaizApellaniz2004,Sota2008,Handler2012,Williams+2013,Cazorla+2017,PozoNunez2019,TriguerosPaez2021,Johnston2021}
and transiting exoplanets observations
\citep[e.g.,][]{Henry2000,Charbonneau2000,Mandel2002,Seager2003}. The
observed radius of a companion star/exoplanet obtained through these
methods rely on the assumptions in radius of the primary/host star
(e.g., \citealt{Johnson+17}). Our method of analysis could be used to break down the radius uncertainty budget in any of these situations, or for other applications involving stars of different masses.

\subsection{Other Effects on Stellar Radii} \label{sec:other_effects}

The initial hydrogen mass fraction $X$ and stellar wind-metallicity $Z_{\rm wind}$ were held fixed in all of the models presented thus far (one exception being the value of $X$ in the fiducial $2Z$ model), in order to isolate effects that arise from the metallicity-dependent microphysics from those related to the associated change in the hydrogen and helium abundances, or the details of the wind mass-loss prescription.  This section attempts to quantify the impact of these specific choices on stellar radii relative to the microphysics-effects of interest.  We conclude by describing some technical caveats associated with our calculations that also contribute to radii uncertainties.

\paragraph*{Wind Metallicity Z$_{\rm wind}$.}

\begin{figure}
    \centering
    \includegraphics[width=\columnwidth]{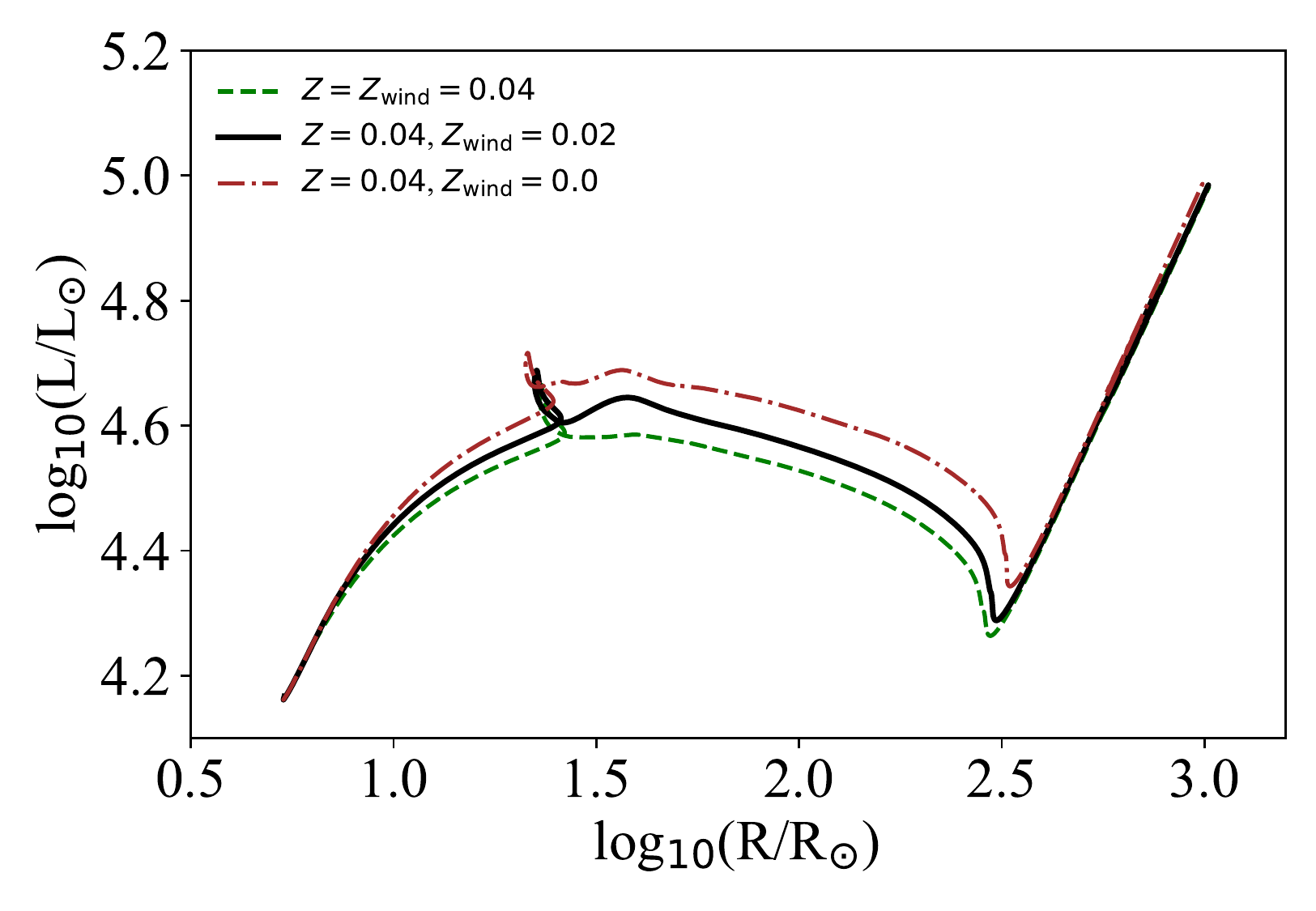}
    \caption{Luminosity-radius tracks for three different choices of the metallicity entering the stellar wind prescription -- $Z_{\rm wind}=0.04$ (dashed
      red), $0.02$ (black; fiducial assumption of all high-Z models in Table \ref{tab:models}) and 0 (dot-dashed purple, no mass loss). All models assume $X=0.75$, $Y=0.21$.}
    \label{fig:hr_fiducial_tests}
\end{figure}

Figure \ref{fig:hr_fiducial_tests} shows how the luminosity-radius evolution of the $Z = 0.04$ fiducial models change for different assumptions about the wind metallicity.  As $Z_{\rm wind}$ increases from 0 to 0.04, the star loses mass at a higher rate, causing a reduction in its luminosity.  The impact of $Z_{\rm wind}$ on the stellar radius at the key epochs of interest are summarized in Table \ref{tab:effect_of_wind}.

In the high-Z models, increasing the value of $Z_{\rm wind}$ from $0$
to $0.04$ results in only a small change in the radius
($\Delta_{\rm 2Z} \lsim 5$\%), except during the Hayashi epochs when
the change is as large as $\Delta_{\rm 2Z} \sim 10\%$. For comparison,
the largest single microphysics-dependent change during the Hayashi
phase was $\Delta_{\kappa}\approx 8-12$\%. Thus, the impact of the
metallicity dependence of microphysics on stellar radii is at most
comparable to the impact of the metallicity dependence of the
mass-loss prescription ($\dot{M}\propto Z^{0.85}$, \citealt{Vink2005}), at least as can be encapsulated through
order-unity changes in the parameter $Z_{\rm wind}$.

Conversely, the low-Z models can exhibit larger variations by changing
$Z_\mathrm{wind}$ compared to changing the metallicity of
microphysical ingredients.
Increasing $Z_{\rm wind}$ from 0 to 0.04 causes $\Delta_{2Z}$ to
change by as much as $\sim50-70$\%, however the large variations
occur during HG and He core burning phases, where the numerical
uncertainties are also large (Appendix \ref{appendix:numerical}).
Besides these epochs, on the MS and the Hayashi track, the $Z_{\rm wind}$-dependence is smaller than the dominant microphysics effects.  Namely, $\Delta_{\epsilon}=5$\% on the MS, compared to $\sim0.3-1$\% changes due to $Z_{\rm wind},$ and $\Delta_{\kappa}=1-6$\% on the Hayashi track, compared to $\lsim1$\% due to $Z_{\rm wind}$.

In addition to changes in stellar radii that arise from changing $Z_{\rm wind}$, different choices about the wind prescription itself have an impact.  \cite{renzo:17} explored the uncertainties in stellar radii that arise from imposing different wind mass-loss prescriptions.  They find that using different wind parameters results in $\sim$7\% change in stellar radius in the RSG phase, for example.

\begin{table}
\centering
\begin{tabular}{c|c|c|c|c|c|c}
    & \multicolumn{3}{c|}{High-Z} & \multicolumn{3}{c}{Low-Z} \\
    \cline{1-4} \cline{5-7}
    $Z_{\rm wind}$ & 0.04 & 0.02$^{\star}$& 0.0 & $2\times10^{-3}$& ${10^{-3}}^{\star}$ & 0.0 \\
    \hline
    MS & 27.0 & 24.3 & 19.1 & 8.8 & 8.8 & 9.1\\
    HG & -7.6 & -5.8 & 0.1 & -38 & 26.9 & -43.7 \\
    He Burning & 3.6 & 4.7 & -2.5 & -3.5 & 51.3 & 12.0\\
    Early Hayashi & -2.4 & -8.3 & -6.0 & -0.1 & 1.1 & 0.45 \\
    End Hayashi & 13.6 & 12.6 & 8.6 & 5.9 & 6.8 & 5.7 \\
    \hline
\end{tabular}
\caption{Percentage change, $\Delta_{\rm 2Z}$, in the radius of the $2Z$ fiducial model relative to the fiducial $Z$ model as a result of adopting the wind metallicity $Z_{\rm wind}$ denoted in each column.  In all other input, the high- and low-Z models assume $Z=0.04$ and $Z=2\times10^{-3}$, respectively.  $^{\star}$Wind metallicity used in the $2Z$ fiducial models in Table \ref{tab:models}.  }
\label{tab:effect_of_wind}
\end{table}

\begin{figure}
    \centering
    \includegraphics[width=\columnwidth]{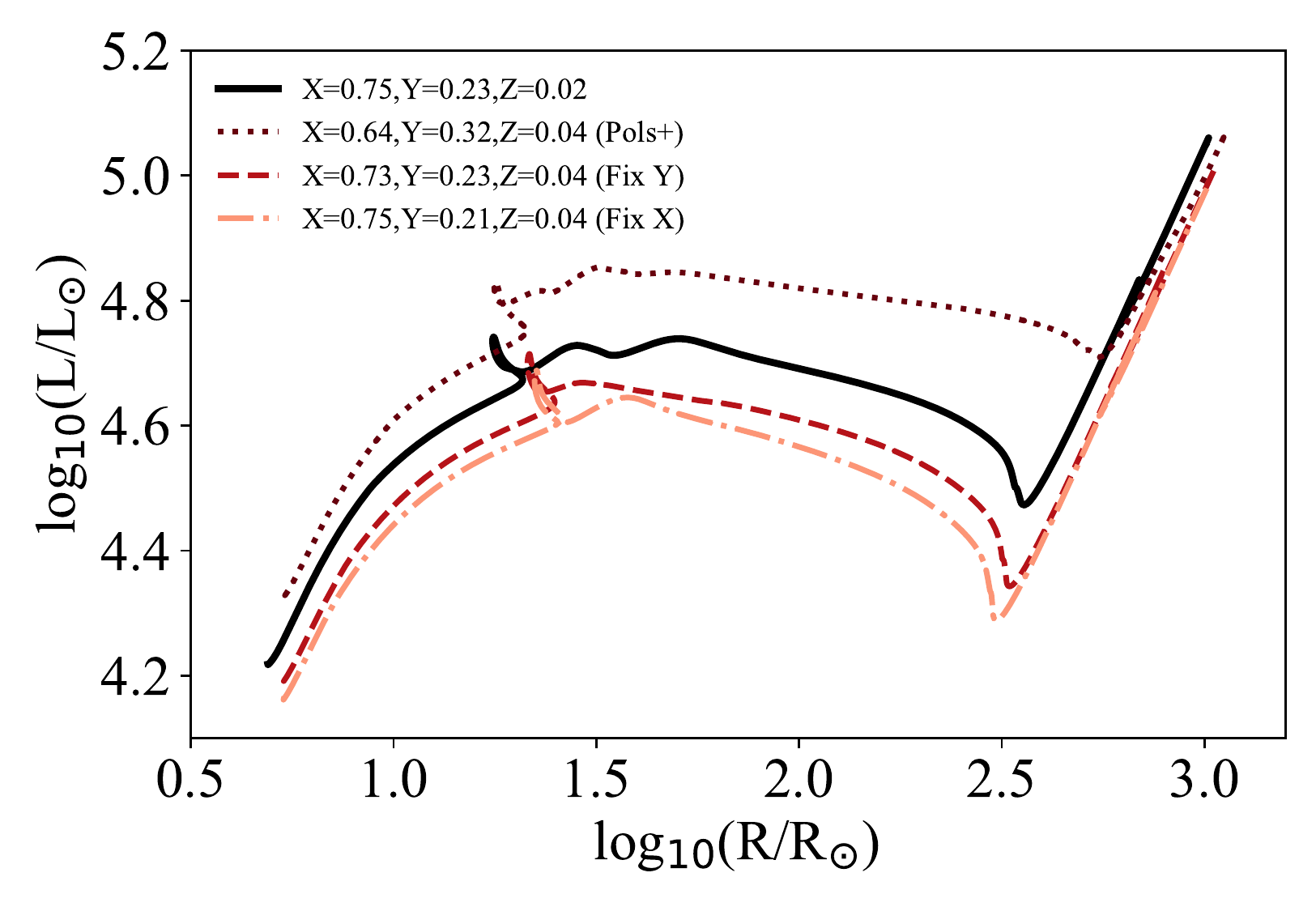}
    \caption{Dependence of the luminosity-radius evolution in the high-Z models on the choice of initial hydrogen $X$ and helium $Y$ mass-fractions. The fiducial $Z=0.02$ model, for which $X=0.75$ and $Y=0.23$, is shown with a black curve.  For the $Z=0.04$ case, we compare three different choices: $X=0.64$, $Y=0.32$, i.e. following the default scaling in \texttt{MESA} \citep{Pols+98} (dotted maroon line); $Y=0.23$ is fixed and $X=1-Y-Z=0.75$ decreases (red dashed line); $X$ is fixed at $0.75$ and $Y=1-X-Z=0.21$ decreases (orange dot-dashed line).  This last option ($X$ fixed) corresponds to that adopted in our $2Z$ models in Table \ref{tab:models}.}
    \label{fig:hr_y}
\end{figure}

\paragraph*{Hydrogen and Helium Abundances.}  Figure \ref{fig:hr_y} compares the effects of different choices for the initial hydrogen ($X$) or helium ($Y$) mass fractions on the luminosity-radius evolution of the high-Z fiducial models ($Z=0.02$ and $Z=0.04$).  We show models in which $X$ or $Y$ are held fixed with varying $Z$, as well as those where $X, Y$ vary together following \citet{Pols+98}.  Table \ref{tab:effect_of_Y} summarizes the fractional change in the stellar radii $\Delta_{2Z}$ for these choices at the key epochs of evolution.

The effect of changing $Y$ illustrated in Figure~\ref{fig:hr_y} broadly agrees with those found by \citet{Farrell+21} through a different numerical experiment.  Specifically, fixing $Y$ changes the value of $\Delta_{2Z}$ from the $X-$fixed case by $\sim 1-7$\% and $\lesssim20$\% in high-$Z$ and low-$Z$ models, respectively (Table \ref{tab:effect_of_Y}); these variations are smaller than the dominating effect of microphysics at any given epoch, except at the HG and He core burning where large numerical uncertainties persist.  Varying $X$ and $Y$ together following \citet{Pols+98} changes $\Delta_{2Z}$ by an even larger amount compared to the $X-$fixed case, since now neither the hydrogen nor helium content remain the same as in the fiducial $2Z$ models.

The different value of $Y$ which characterize our $\{Z_{\kappa}, Z_{\mu}, Z_{\epsilon}\} = 2Z$ models versus that of the fiducial $2Z$ model to which they are compared, introduces another source of uncertainty in attributing radius changes solely to individual $Z$-dependent microphysical effects, in addition to those described in Appendix \ref{appendix:numerical}.

\begin{table}
\centering
\begin{tabular}{c|c|c|c|c|c|c}
    & \multicolumn{3}{c|}{High-Z} & \multicolumn{3}{c}{Low-Z} \\
    \cline{1-4} \cline{5-7}
    Composition & Pols+& Fix Y & Fix X$^{\star}$ & Pols+& Fix Y & Fix X$^{\star}$\\
    \hline
    MS & 1.9 & 20.0 & 24.3 & 9.4 & 8.7 & 8.8 \\
    HG & 0.1 & -0.5 & -5.8 & 9.7 & 15.2 & 26.9 \\
    He Burning & 9.4 & 10.1 & 4.7 & 10.0 & 27.6 & 51.3 \\
    Early Hayashi & 2.3 & -0.7 & -8.3 & 0.03 & 0.6 & 1.1 \\
    Late Hayashi & 8.0 & 11.8 & 12.6 & 6.9 & 6.9 & 6.8 \\
    \hline
\end{tabular}
\caption{Percentage fraction change, $\Delta_{\rm 2Z}$, in the radius of the $2Z$ fiducial model relative to the fiducial $Z$ model as a result of varying the initial hydrogen $X$ and helium $Y$ mass fractions at different
  epochs of stellar evolution as marked. From left to right, the columns assume initial composition values (X,Y,Z) as follows: $(0.64,0.32,0.04)$, $(0.73,0.23,0.04)$,
  $(0.75,0.21,0.04)$, $(0.754,0.244,2\times10^{-3})$,
  $(0.749,0.249,2\times10^{-3})$, and
  $(0.75,0.248,2\times10^{-3})$.  In all other input, the high- and low-Z models assume $Z=0.04$ and $Z=2\times10^{-3}$, respectively.  $^{\star}$Composition used in the $2Z$ fiducial models introduced in Section \ref{sec:method}. }
\label{tab:effect_of_Y}
\end{table}

    \paragraph*{Limitations of 1D modeling.} The post-MS evolution of massive stars is notoriously
  sensitive to the composition and entropy profiles between the H
  depleted core and the envelope \citep[e.g.,][]{walmswell:15,schootemeijer:19}. 
  These aspects are determined by processes hard to model in 1D, such as the recession of the convective boundary during the
  MS, semiconvective and rotational mixing above the core during and
  after H-core burning \citep[e.g.,][]{langer:89, maeder:2000, schootemeijer:2018, schootemeijer:19, klencki:2020a, klencki:2021}, the
  mass-loss history \citep[e.g.,][]{belczynski:08,renzo:17}, and the
  possible accretion episodes \citep[e.g.,][]{Renzo2021}, all of
  which can have different metallicity dependencies.
  For stars more massive than the $15\,M_\odot$ models that we
    considered here, another possibly metallicity-dependent,
    macrophysical ingredient enters in the radius determination:
    the treatment of energy transport in radiatively
    inefficient super-Eddington convective layers
    \citep[e.g.][]{Paxton+2013, jiang:15, jiang:18}. The detailed
    treatment of such regime, which is an intrinsically
    three-dimensional problem because of the interplay between
    turbulence seeded at the iron opacity bump and the helium opacity
    bump \citep{jiang:18}, determines whether more massive star
    inflate at the end of their main sequence \citep{sanyal:15, sanyal:17} or instead
    trigger eruptive mass loss events. However, our 15$\,M_\odot$
    models are sufficiently low-luminosity to not enter this regime
    common for $M\gtrsim 25\,M_\odot$.
  \paragraph*{Comparison with observational uncertainty.}
    Nevertheless, we are being extremely demanding with the
    numerical resolution adopted in our models.  Appendix \ref{sec:resolution} shows
    that the largest uncertainties that arise from varying the
    spatial and time resolution parameters occur on the Hertzsprung
    Gap and has magnitude of $\lesssim10$\%.  In reality, the
    observational errors on the stellar radii of massive stars are
    much larger compared to the variations we see here.  For example, the closest O-type MS star, $\zeta$ Ophiuchi ($\sim20M_{\odot}$), has a radius uncertainty of
$\sim10-20$\% ~\citep[see, e.g.,][and references therein]{villamariz:05, gordon:18, Renzo2021,
shepard:22}, while the radius uncertainty of the nearest RSG $\alpha$
Orionis (Betelgeuse, $\sim15-20M_{\odot}$) is up to $\sim30$\%
\citep[e.g.,][and references therein]{Chatzopoulos+2020}.

\section{Conclusions}
\label{sec:conclusion}

\begin{figure*}
    \centering
    \includegraphics[width=\columnwidth]{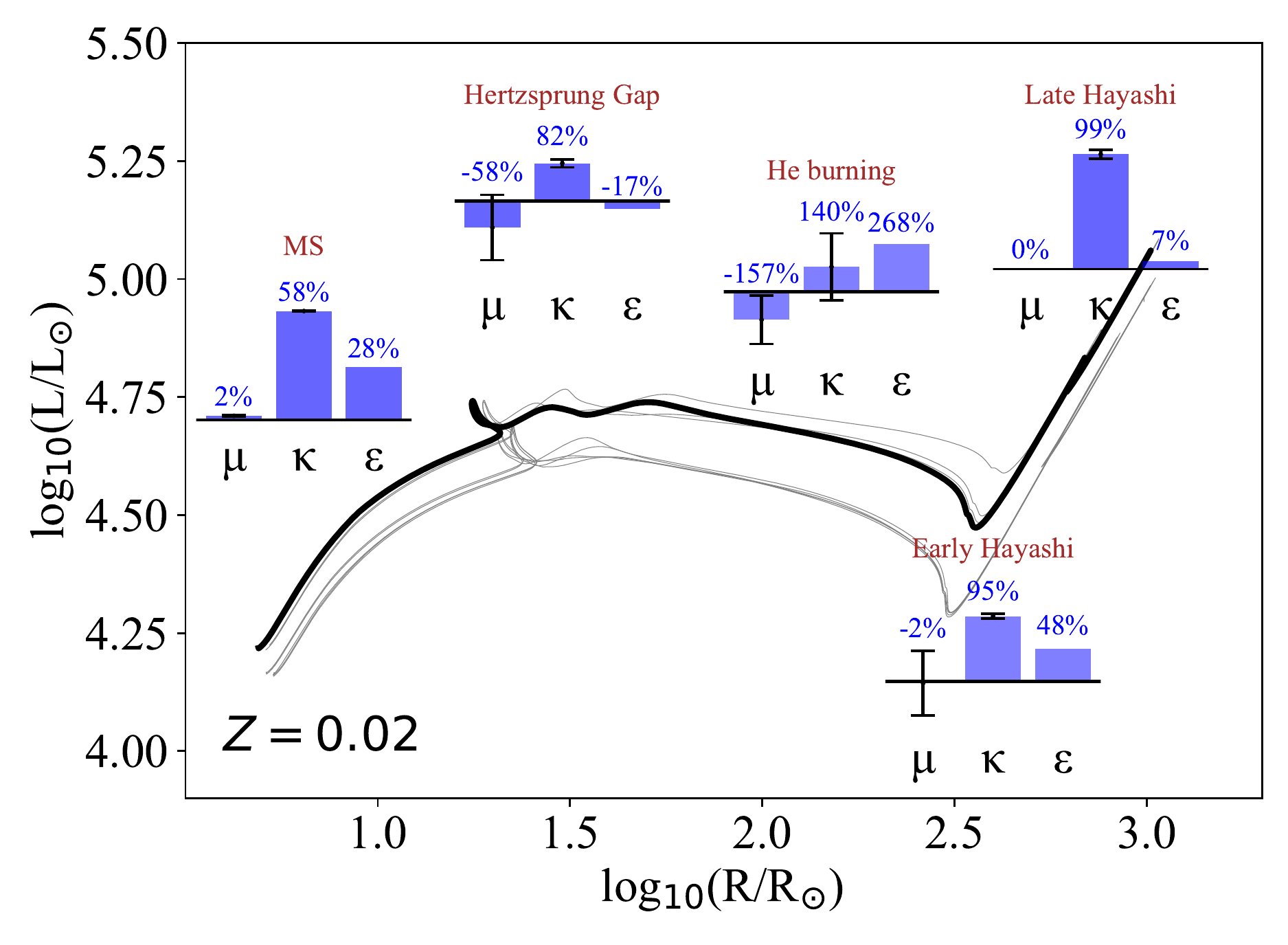}
    \includegraphics[width=\columnwidth]{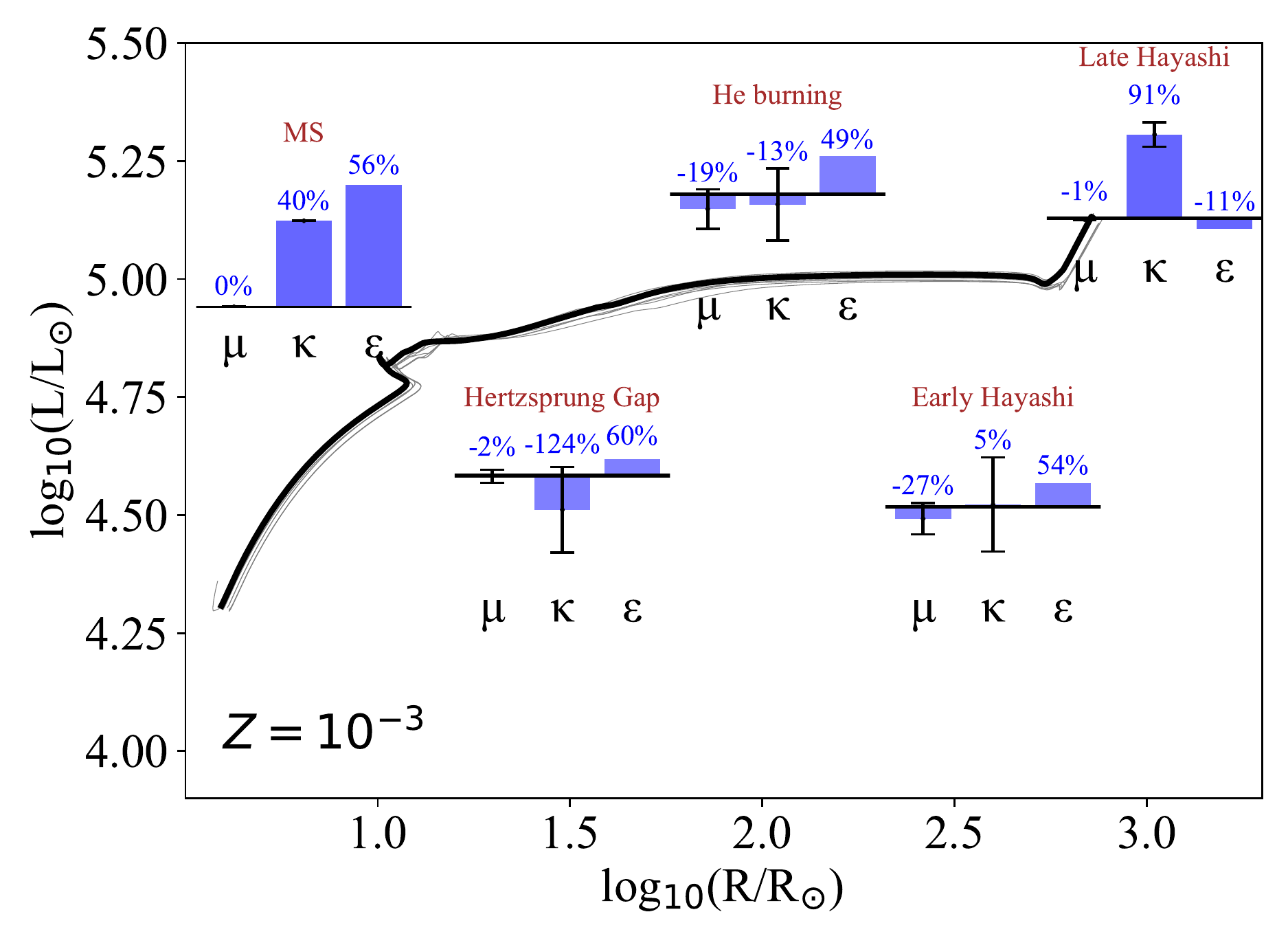}
    \caption{Partitioning of the effects of microphysics on stellar
      radii at different epochs of stellar evolution in massive stars at approximately solar (left) and sub-solar (right) metallicity.  Black curves are the luminosity-radius tracks of the fiducial $Z=0.02$ and $Z=10^{-3}$ models, while the light gray tracks are the variable-microphysics models from Fig~\ref{fig:HR_4panel}.  Bar charts next to each key epoch show the relative effect of the stellar radius of each microphysics input, $\mu$, $\kappa$ and $\epsilon$.  The height of the bars represents the fraction (in percent) obtained from the ratio of the $\Delta$-element corresponding to doubling each microphysics metallicity over the full change $\Delta_{2Z}$, while the error bars are obtained from the corresponding null matrix (Eqs.~\ref{eq:bar}, \ref{eq:err_bar}).  In the solar-$Z$ case, opacity has the largest impact on stellar
    radius at most stages of the evolution despite the uncertainties,
    except during He core burning.  In the subsolar case, nuclear
    physics has the largest impact on radius at most epochs,
    although during some phases (Hertzsprung Gap, He core burning, and the Early Hayashi phase) the uncertainties are larger than the physical effect we are trying to measure so no firm conclusions can be drawn.}
    \label{fig:bar_plot}
\end{figure*}

We have constructed \texttt{MESA} stellar evolution models of a
15M$_{\odot}$ star to dissect the effects
that metallicity-dependent opacity, mean molecular weight and nuclear
physics on stellar radii at key points in the star's life.  We explore models centered around a solar ($Z=0.02$)
and sub-solar ($Z=10^{-3}$) metallicity value, which characterize the isolated and aggregate impacts of the three $Z$-dependent
microphysics.

As a visual demonstration of the fractional radius-change matrices presented in Sec.~\ref{sec:results}, we summarize our findings in
Fig.~\ref{fig:bar_plot}.  We show the partitioning of microphysics with
bar plots for each epoch in the high-Z and low-Z case. The heights and error bars are defined as: 
\begin{equation} \label{eq:bar}
	{\rm (Height \ of \ bar)} = \frac{\Delta_i}{\Delta_{2Z}} \times 100\% \ \
\end{equation}
\begin{equation} \label{eq:err_bar}
	{\rm (Error \ bar)} = \frac{\delta_i}{\Delta_{2Z}} \times 100\% \ \
      \end{equation}
$i\in \{\mu,\kappa,\epsilon\}$ corresponds to each diagonal value in the matrix.
The error bars are obtained from the null matrices (`$\delta$'s') given in Appendix~\ref{sec:appendix_a}, which represent the uncertainties that arise from our
implementation method. 
Our main findings are summarized as follows:

\begin{itemize}
	\item In the high-Z models, $Z$-dependent opacity is the dominating effect of microphysics on radius in all except the helium core burning epoch of the evolution. However the uncertainties at this epoch is too large, which prevents us from drawing definitive conclusion.
	\item In the low-Z models, nuclear
          reactions that depends on $Z_{\epsilon}$ is the dominating
          microphysics on stellar radius on the MS, while
          opacity becomes the most important in late Hayashi track,
          where the uncertainties at these epochs are sufficiently
          small ($\delta$ < $\Delta$).
	\item During the 
          intermediate evolutionary stages at low-Z
          -- He core burning, Hertzsprung Gap and early Hayashi track -- opacity is the biggest effect, if we take into account the uncertainties in the matrices.  Had we neglected the numerical uncertainties, opacity would be the dominating effect at Hertzsprung Gap, nuclear reactions at He burning and early Hayashi track.
	\item In both high-Z and low-Z cases, the $Z$-dependent effect of EOS is the weakest on radius across all five epochs.
	\item Our numerical results on the MS and at late Hayashi track broadly agree with the corresponding analytical solutions, under reasonable assumptions.
	\item Our choice to fix stellar wind metallicity to
          $Z_{\rm wind}=Z$ in high-Z and low-Z models and to fix
          $X=0.75$ in all models, does not change the resulting
          partitioning of the microphysics. The different
          assumptions about $Z_{\rm wind}$ and the composition ($X$
          and $Y$) influence the $2Z$ fiducial models, although the
          influence is small compared to that of the microphysics during most epochs.

	\item Our results highlight the importance of quantifying uncertainties in stellar evolution models, particularly when conducting numerical experiments where the measurable effects can be subtle.  In this work, we evaluate the uncertainties resulting from (i) our specific implementations of different microphysical metallicities in \texttt{MESA}, (ii) the adopted time and spatial resolution.  We conclude that our implementation leads to slightly larger uncertainties than numerical resolutions.
\end{itemize}

Overall our findings support the promise of controlled numerical experimenters with stellar evolution codes to elucidate the dominant physical processes at work.
Future work could include the application of our method to stars of other masses, in order to systematically dissect the effects of different microphysics on macrophysical observables, such as stellar age, luminosity, rotation, stellar mass-loss, etc..  The matrix formalism we have introduced provides a convenient framework to present and interpret these results.

\vspace{\baselineskip}

\appendix

\section{Numerical Uncertainties}
\label{appendix:numerical}

\begin{figure*}
    \centering
    \includegraphics[width=2\columnwidth]{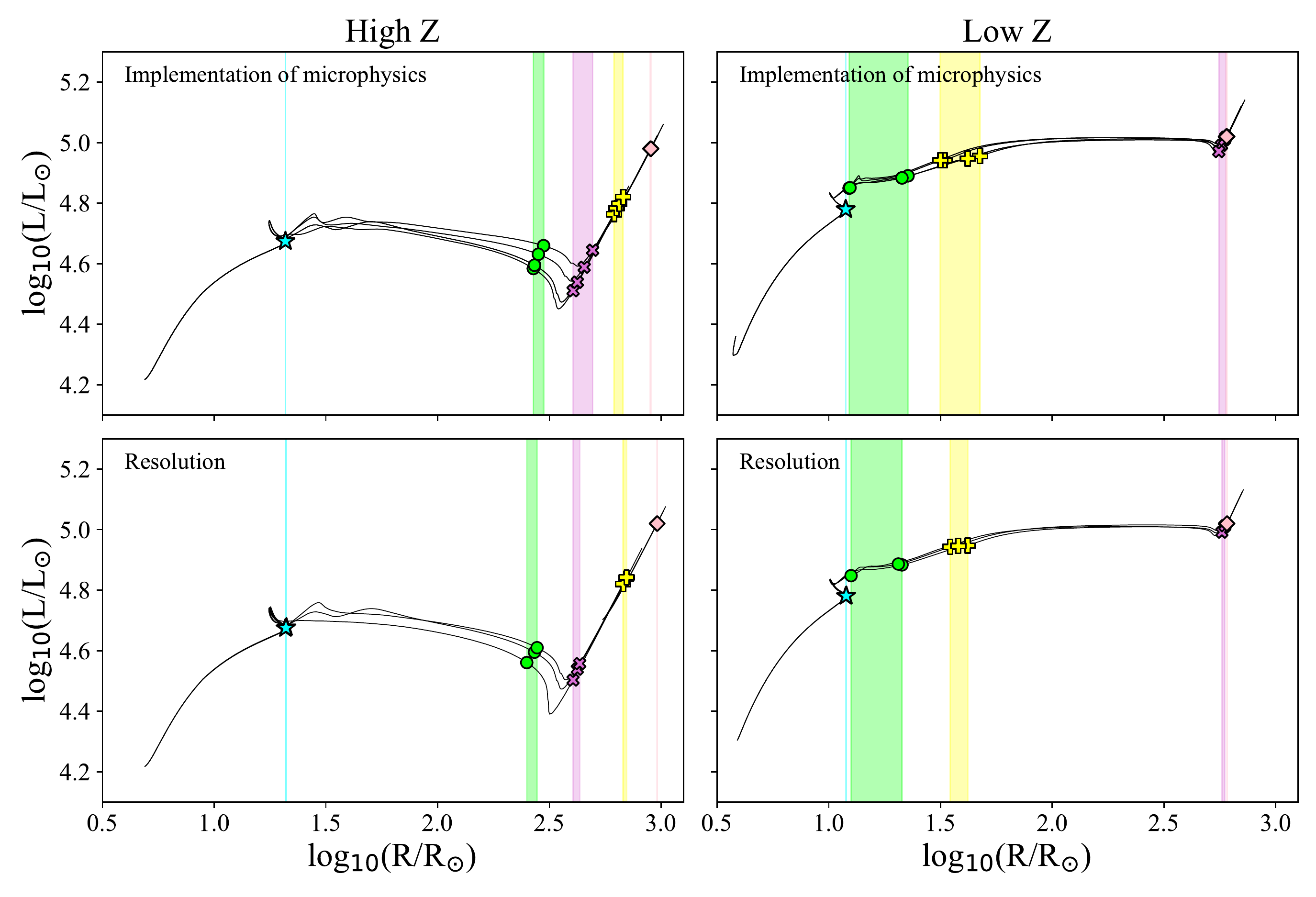}
    \caption{Evolution of stellar luminosity with radius from a grid of ``null'' models used to assess the numerical uncertainties in our results due to (1) our implementation of the metallicity-dependent microphysics in \texttt{MESA} (top panels; Sec.~\ref{sec:appendix_a}); and (2) temporal and mass-grid resolution (bottom panels; Sec.~\ref{sec:resolution}).  We show separately the high-$Z$ ($Z = 0.02$; left panels) and low-$Z$ ($Z = 10^{-3}$; right panels) models.
    The numerical implementation models were run with $Z_{\kappa,\mu,\epsilon}=Z_{\rm wind} = Z$ for all different combinations of $\{\kappa,\mu,\epsilon\}$.  For the grid resolution runs, the time
      resolution parameter ``vct" was varied between $10^{-3}$ to $10^{-4}$ and
      mass-coordinate resolution $d_x$ from 1.0 to 0.5.  The five key evolutionary epochs are marked with the same symbols as those in Fig. \ref{fig:HR_4panel}. The spread of the points in
      radius-coordinate in each epoch indicates the degree of
      uncertainty -- highlighted with colored bands. Comparing the left and right panels across each row, we see that the uncertainties in the high-Z models are small compared to those in the low-Z case.  A comparison between the top and
      bottom panels in each column shows that the uncertainty introduced by our implementation is typically comparable or larger than that arising from our choice of numerical resolution.}
    \label{fig:appendix}
\end{figure*}

Here we evalulate the numerical and algorithmic
uncertainties associated with the method introduced in Sec.~\ref{sec:method}.
These uncertainties arise from: (1) our method of implementing $Z$-dependent
microphysics into \texttt{MESA}; (2) numerical resolution, both in time and in mass-coordinate.  The amplitudes of these two effects are summarized and compared in Fig. \ref{fig:appendix}.

\subsection{Microphysics Implementation} \label{sec:appendix_a}

We first explore the impact of the artificial modifications we have made to the microphysics input to \texttt{MESA} on the precision of our results for stellar radii.  As a sanity check, we rerun all the models in Table~\ref{tab:models} using our customized routines, except now fixing the input metallicity $Z$ at its original value $Z_{\mu}=Z_{\kappa}=Z_{\epsilon} = Z$ rather than doubling it ($Z_{\mu}=Z_{\kappa}=Z_{\epsilon} = 2Z$) as before.  Models \#1 and \#9
are now assigned the same metallicity ($Z = 0.02$), as are models \#10
and \#18 ($Z = 10^{-3}$); however, the altered wind-metallicity is only
activated in models \#9 and \#18, both at $Z_{\rm wind} = Z$.  We refer to the new series of radius matrices generated by comparing the $Z_{\mu}=Z_{\kappa}=Z_{\epsilon} = Z$ and fiducial $Z$ models as ``null'' because, given a perfectly controlled experiment,
their entries should all be strictly zero.  We denote the elements of
the null matrices by $\delta$'s to distinguish them from the normal matrix elements $\Delta$'s
introduced in \S\ref{sec:analysis}.

The expectation $\delta \rightarrow 0$ will not be realized in practice if our alterations to the microphysics input of \texttt{MESA} (\S~\ref{sec:microphysics}) do not exactly reproduce the default assumptions present in the fiducial $Z$ model.  The magnitude of the residual $\delta$ values can therefore be used to estimate the minimum uncertainty on the $\Delta$ values presented in $\S\ref{sec:results}$ (in addition to the effects of resolution explored in Appendix \ref{sec:resolution}).

The null matrices corresponding to each phase of stellar evolution are
given below.  The value of $\delta_{\epsilon}$ is always
zero; this is because our alterations to the nuclear reactions input
involve simply changing the reaction rate factors, which is a built-in
option already present in \texttt{MESA}. Also note that $\delta_Z$ is small
compared to the other $\delta$'s in all cases; this matrix entry compares two fiducial models with the same $Z$ but different $Z_{\rm wind}$. The fact that $\delta_Z$ is small (strictly zero in the case of MS and late Hayashi track) shows that
our method of artificially fixing the value $Z_{\rm wind}$ in our numerical experiments does
not itself impart a significant uncertainty to $\Delta$ (however, see discussion in Sec.~\ref{sec:other_effects}).

In general the other $\delta$ matrix elements are non-zero. The radii
at the five key evolutionary points discussed in Sec \ref{sec:results} are
marked as symbols in Fig. \ref{fig:appendix}. The spread in radii
between models is depicted graphically by the colored bands, enabling
an easy comparison of the uncertainty that arise due to our microphysics
implementation (top panels) and due to the numerical resolution
(bottom panels; Sec.~\ref{sec:resolution}).

In broad terms, the implied uncertainties in the high-metallicity
$Z = 0.02$ model sequence (top panel left of Fig. \ref{fig:appendix})
are sufficiently small so as to not dominate the physical radii changes we are trying to measure (the $\delta$'s are much smaller than
the $\Delta$'s), except during the He core burning and early Hayashi phases.  However, in the $Z = 10^{-3}$ models (top right panel of Fig. \ref{fig:appendix}), the algorithmic
uncertainties become large also in the Hertzsprung Gap, and are generally larger at all epochs than for the high-Z models.

Below now present and describe the null matrices for each stage of
stellar evolution in greater detail.

\begin{center}
{\bf Main Sequence}

{\bf High $Z$}

\begin{blockarray}{ccccl} \label{mat:format}
   $\delta$ (\%) & $\mu$ & $\kappa$ & $\epsilon$ & \\
\begin{block}{c[ccc]l}
  $\mu$ & 0.072 & 0.127 & 0.072 & \ \ $\delta_{\mu\kappa\epsilon} = 0.127 $\\
  $\kappa$ & - & 0.074 & 0.074 & \ \ Tr$(\delta)= 0.074 $ \\
  $\epsilon$ & - & - & 0.0 & \ \ $\delta_{Z}=0.0$\\
\end{block}
\end{blockarray} \\

\newpage
{\bf Low $Z$}

\begin{blockarray}{ccccl} \label{mat:format}
   $\delta$ (\%) & $\mu$ & $\kappa$ & $\epsilon$ & \\
\begin{block}{c[ccc]l}
  $\mu$ & 0.026 & 0.017 & 0.026 & \ \ $\delta_{\mu\kappa\epsilon} = 0.017$\\
  $\kappa$ & - & 0.006 & 0.006 & \ \ Tr$(\delta) = 0.032$ \\
  $\epsilon$ & - & - & 0.0 & \ \ $\delta_{Z}=0.0$\\
\end{block}
\end{blockarray} \\
\end{center}
On the MS, the null matrices are small $\delta (\lesssim 1\%)$ relative to the $\Delta$ values in both high-Z and low-Z models.  Our results on the MS (\S\ref{sec:MS}) are the most robust of all the epochs we have explored.

\begin{center}

{\bf Hertzsprung Gap}

{\bf High $Z$}

\begin{blockarray}{ccccl} \label{mat:format}
   $\delta$ (\%) & $\mu$ & $\kappa$ & $\epsilon$ & \\
\begin{block}{c[ccc]l}
  $\mu$ & 9.9 & 4.1 & 9.9 & \ \ $\delta_{\mu\kappa\epsilon} = 4.1 $\\
  $\kappa$ & - & -1.2 & -1.2 & \ \ Tr$(\delta) = 8.7 $ \\
  $\epsilon$ & - & - & 0.0 & \ \ $\delta_{Z}=-3.4$\\
\end{block}
\end{blockarray} \\
\newpage
{\bf Low $Z$}

\begin{blockarray}{ccccl} \label{mat:format}
   $\delta$ (\%) & $\mu$ & $\kappa$ & $\epsilon$ & \\
\begin{block}{c[ccc]l}
  $\mu$ & 6.4 & -42 & 6.4 & \ \ $\delta_{\mu\kappa\epsilon} = -42$\\
  $\kappa$ & - & -42 & -42 & \ \ Tr$(\delta) = -35$ \\
  $\epsilon$ & - & - & 0.0 & \ \ $\delta_{Z}=-1.7$\\
\end{block}
\end{blockarray} \\
\end{center}
In the Hertzsprung Gap, the null matrix entries are sufficiently
small in the high-Z models relative to the $\Delta$'s, that our
conclusion from Sec.~\ref{sec:HG} that metallicity entering the opacity
has the greatest impact on stellar radii at this stage still holds
well. By contrast, in the low-Z models the $\delta$ entries are large
compared to the corresponding $\Delta$ values, precluding definitive
conclusions about the dominant microphysical effect at this epoch.

\begin{center}
{\bf Helium Core burning}

{\bf High $Z$}

\begin{blockarray}{ccccl} \label{mat:format}
   $\delta$ (\%) & $\mu$ & $\kappa$ & $\epsilon$ & \\
\begin{block}{c[ccc]l}
  $\mu$ & -6.4 & -3.9 & -6.4 & \ \ $\delta_{\mu\kappa\epsilon} = -3.9 $\\
  $\kappa$ & - & -8.9 & -8.9 & \ \ Tr$(\delta) = -8.90 $ \\
  $\epsilon$ & - & - & 0.0 & \ \ $\delta_{Z}=0.9$\\
\end{block}
\end{blockarray} \\
\newpage
{\bf Low $Z$}

\begin{blockarray}{ccccl} \label{mat:format}
   $\delta$ (\%) & $\mu$ & $\kappa$ & $\epsilon$ & \\
\begin{block}{c[ccc]l}
  $\mu$ & 13.4 & -21.2 & 13.4 & \ \ $\delta_{\mu\kappa\epsilon} = -21.2$\\
  $\kappa$ & - & -24.6 & -24.6 & \ \ Tr$(\delta) = -11.2$ \\
  $\epsilon$ & - & - & 0.0 & \ \ $\delta_{Z}=-5.7$\\
\end{block}
\end{blockarray} \\
\end{center}
Similarly, at He core burning the uncertainties implied by the
null-matrix values are too large compared to the $\Delta$ values
presented in Sec.~\ref{sec:he_ignition}. In the high-Z models,
$\delta_{\kappa}$ and $\delta_{\kappa\epsilon}$ are large compared to
$\Delta_{\kappa}$ and $\Delta_{\kappa\epsilon}$, challenging our
conclusion that nuclear reactions are the dominant effect on stellar
radii at this epoch (even though
$\Delta_{\epsilon}\gg\Delta_{\kappa}$). In the low-Z models, the
$\delta$-implied uncertainties are also too large, again preventing us
from drawing meaningful conclusions.

\begin{center}

{\bf Early Hayashi track}

{\bf High $Z$}

\begin{blockarray}{ccccl} \label{mat:format}
   $\delta$ (\%) & $\mu$ & $\kappa$ & $\epsilon$ & \\
\begin{block}{c[ccc]l}
  $\mu$ & 3.9 & 1.3 & 3.9 & \ \ $\delta_{\mu\kappa\epsilon} = 1.3 $\\
  $\kappa$ & - & -0.3 & -0.3 & \ \ Tr$(\delta) = 3.6 $ \\
  $\epsilon$ & - & - & 0.0 & \ \ $\delta_{Z}=1.0$\\
\end{block}
\end{blockarray} \\
{\bf Low $Z$}

\begin{blockarray}{ccccl} \label{mat:format}
   $\delta$ (\%) & $\mu$ & $\kappa$ & $\epsilon$ & \\
\begin{block}{c[ccc]l}
  $\mu$ & 0.35 & -0.73 & 0.35 & \ \ $\delta_{\mu\kappa\epsilon} = -0.73$\\
  $\kappa$ & - & -1.2 & -1.2 & \ \ Tr$(\delta) = -0.8$ \\
  $\epsilon$ & - & - & 0.0 & \ \ $\delta_{Z}=-0.1$\\
\end{block}
\end{blockarray} \\

\end{center}
The envelope mass becomes 50\% convective at the beginning of the
Hayashi track, as occurs before (in the high-Z models) or after (in
low-Z models) helium core burning. At this epoch, the effect of
doubling the metallicity values in the low-Z models is small compared
to those in high-Z models (see Fig.~\ref{fig:HR_4panel}). The
magnitudes of the $\delta$'s in the low-Z null matrix are small but
nevertheless greatly exceed the corresponding $\Delta$ values (Sec.~\ref{sec:hayashi_1}). In the high-Z models, $\delta_{\mu}$ and
$\delta_{\mu\epsilon}$ are large compared to $\Delta_{\mu}$ and
$\Delta_{\mu\epsilon}$ at this epoch, but these do not dominate the
radius-dependence at this epoch. Our main conclusions in the high-Z
case are therefore robust to the numerical uncertainties explored in
this section.

\begin{center}
{\bf Late Hayashi track}

{\bf High $Z$}

\begin{blockarray}{ccccl} \label{mat:format}
   $\delta$ (\%) & $\mu$ & $\kappa$ & $\epsilon$ & \\
\begin{block}{c[ccc]l}
  $\mu$ & 0.03 & 0.25 & 0.03 & \ \ $\delta_{\mu\kappa\epsilon} = 0.25 $\\
  $\kappa$ & - & -0.48 & -0.48 & \ \ Tr$(\delta)= -0.48 $ \\
  $\epsilon$ & - & - & 0.0 & \ \ $\delta_{Z}=0.0$\\
\end{block}
\end{blockarray} \\
{\bf Low $Z$}

\begin{blockarray}{ccccl} \label{mat:format}
   $\delta$ (\%) & $\mu$ & $\kappa$ & $\epsilon$ & \\
\begin{block}{c[ccc]l}
  $\mu$ & -0.06 & -0.84 & -0.06 & \ \ $\delta_{\mu\kappa\epsilon} = -0.84$\\
  $\kappa$ & - & -0.92 & -0.92 & \ \ Tr$(\delta) = -0.99$ \\
  $\epsilon$ & - & - & 0.0 & \ \ $\delta_{Z}=0.0$\\
\end{block}
\end{blockarray} \\

\end{center}
The late Hayashi track matrices exhibit the second best accuracy (lowest $\delta$ values), next to those of the MS, in both high-Z and low-Z model sequences. The uncertainties are sufficiently small compared to all of the $\Delta$ values (\S\ref{sec:hayashi_2}) that our conclusions regarding this phase should also be robust.

\subsection{Numerical Resolution} \label{sec:resolution}

Our numerical implementation not only modifies the physics but also numerical details of the \texttt{MESA} models wherever we change the microphysics, including, for example, the precision of the initial abundances that we manually enter for the opacity routine. Such change could result in different resolution requirements in each model. We estimate the uncertainties from choices of numerical resolutions below, and compare to those from our implementation method.

The bottom panels of Fig.~\ref{fig:appendix} show $2Z$ fiducial models
in which the temporal or spatial resolution have been changed from
their default assumptions. Specifically, we vary the two resolutions
separately by changing\footnote{For the time resolution, this is not
  the recommended approach in more recent \texttt{MESA} versions, see
  the \href{https://docs.mesastar.org/en/release-r22.05.1/using_mesa/best_practices.html?highlight=time\%20resolution\#experiment-with-the-temporal-resolution}{documentation}.}
\texttt{mesh\_delta\_coeff} ($d_X=1.0$ and 0.5) and
\texttt{varcontrol\_target} (vct=$10^{-3}$ and $10^{-4}$),
respectively, where smaller values impose more stringent limits on the
mesh and timestep. As a result, the maximum timesteps (dt) decreases
from $10^{5.6}$ to $10^{4.3}$ years and the spatial resolution is
roughly double the default setting in \texttt{MESA}.

Comparing the width of the radii bands at each epoch in Fig.~\ref{fig:appendix}, the uncertainties associated with the choice of resolution in the low-Z models are larger than those in high-Z models.
As a conservative choice, we use the highest set of resolution, \texttt{mesh\_delta\_coeff=0.5} and \texttt{varcontrol\_target=1.0d-4}, in all of our models in this paper.  However if we compare the bottom and the top panels in Fig.~\ref{fig:appendix}, we see that the uncertainty associated with algorithmic implementation of microphysics typically introduces a larger uncertainty than the choice of resolution.

\section*{Acknowledgements}

BDM acknowledges support from the National Science Foundation (grant
AST-2009255). We also thank E. Laplace and M. Cantiello for helpful
feedback on early versions of this manuscript, and A.~Jermyn for comments and help with the EOS.

\section*{Data availability}
The input files, modified MESA routines, and output files are
available on \url{https://doi.org/10.5281/zenodo.6621643}.

\bibliographystyle{mnras}
\bibliography{cx, mr}

\bsp
\label{lastpage}
\end{document}